# Measuring Nanoscale Torques with Cylindrical-Polarization-based Interferometric Scattering Microscopy

Milan Vala[1]*, Ivan Kopal[1,2], Lauren Takiguchi[3,4], Yevhenii Shaidiuk[1], Vítězslav Lužný[1], Łukasz Bujak[1], Pallav Kosuri[3], Marek Piliarik[1]*

[1] *Institute of Photonics and Electronics, Czech Academy of Sciences, Prague, Czech Republic*

[2] *Department of Physical Chemistry, University of Chemistry and Technology, Prague, Czech Republic*

[3] *Salk Institute for Biological Studies, La Jolla, CA, USA*

[4] *Biophysics Program, Stanford University School of Medicine, Stanford, CA, USA*

Correspondence: vala@ufe.cz; piliarik@ufe.cz

## Abstract

The ability to observe rotational dynamics and measure underlying torques is a crucial component in understanding the function and mechanics of nanoscale systems. Yet, direct observation of rotational dynamics at the single-molecule level in liquids remains challenging due to the trade-off between optical detectability and hydrodynamic responsiveness. Labels that are bright enough for rapid readout typically introduce excessive drag, while minimally perturbing probes are difficult to detect at high speed. This limits access to fast rotational dynamics required for direct torque measurements. Here, we introduce cylindrical-polarization-based interferometric scattering microscopy (cypiSCAT), a method encoding the orientation of anisotropic scatterers directly into a single interferometric point spread function, while intrinsically suppressing the isotropic background. We achieve rotational tracking of low-drag orientation labels based on DNA origami-attached gold nanorods with sub-degree angular precision and microsecond temporal resolution, allowing quantitative characterization of nanoscale rotational dynamics. This capability provides direct access to torque metrology at the single-molecule level, here demonstrated through the extraction of optically induced torques as small as ~1 pN nm. Relying on elastic scattering, cypiSCAT combines ultrafast temporal resolution with long observation times, making it well-suited for capturing rapid and rare rotational events and reaction steps in nanoscale biomolecular systems.

## Introduction

Rotation is a fundamental degree of freedom in nanoscale science, where torques govern a wide range of physical and biological processes. The forces and torques acting on molecular constituents are masked by thermal fluctuations, placing extreme demands on sensitivity, angular and temporal resolution, and the non-invasiveness of the probing method. Direct observation of these motions is key to revealing the mechanics of molecular motors[1], nanoscale organization and conformation of biological matter[2], or the operation of synthetic nanomachines[3].

Optical methods have been at the core of efforts to track single-molecule rotation.[4–11] They often rely on an orientation label that transduces the angular position into a measurable optical signal. Reliable detectability within short exposure times requires a large photon count $N$ for a high signal-to-noise ratio (SNR $\propto \sqrt{N}$). This favors larger labels with stronger scattering or fluorescence. Yet, increasing the label size compromises the hydrodynamic responsiveness as the rotational drag $\gamma_{rot}$ scales cubically with the characteristic size of the label $L$ ($\gamma_{rot} \propto L^3$).[1,12] This fundamental trade-off between optical detectability and hydrodynamic properties sets the limits of current approaches.

Single-fluorophore anisotropy methods minimize hydrodynamic perturbation, but are limited to acquisition rates of typically hundreds of frames per second (fps) due to the fundamental cap of the fluorescent photon flux.[13] To boost SNR, larger fluorescent or scattering labels have been employed to observe the rotation of RNA polymerase[14], or rotary ATPases [10,15,16], however, at the cost of responsiveness. The challenge is particularly acute for fast-rotating systems such as DNA-associated motor proteins, which unwind DNA at speeds of up to $10^3$ base pairs per second. [17] Torque measurements in the 10 pN nm range have been reported so far only for a small number of motor proteins using an indirect stalled rotation assay.[18,19] Recently, a multi-fluorophore-labeled DNA-origami construct has enabled transcription tracking under slowed environmental conditions with single-base pair resolution at 200 fps.[7]

Here, we argue that direct molecular torque measurements in liquids require orders of magnitude higher temporal resolution to distinguish torque-driven dynamics from thermal fluctuations (Supplementary Note 14) and to prevent rotational motion blur of highly diffusive, low-drag labels (Supplementary Note 8). In this regard, plasmonic nanorods sized below 100 nm are efficient orientation labels, offering a large scattering cross-section and high anisotropy. Dark-field detection of a 25 nm × 71 nm gold nanorod achieved approximately 4 degrees angular precision with a temporal resolution of 2 microseconds.[8] Differential interference contrast (DIC) tracking of gold nanorods has been demonstrated on membranes[20] and in cells[21], though typically at hundreds of fps. Interferometric scattering microscopy (iSCAT) has emerged as a powerful tool for tracking deeply subwavelength objects[22], detecting single proteins[23–25], and nanoparticles down to 2 nm[26]. Despite its promise, polarization-resolved iSCAT has been reported so far only to measure static nanorod orientations[27,28] or protein-related conformational changes[29,30], leaving the high-speed, rotational tracking with low-drag labels unexplored.

We introduce cylindrical-polarization-based interferometric scattering microscopy, here denoted as cypiSCAT, a technique for microsecond-resolved rotational tracking of nanoscale objects. CypiSCAT extracts the orientation from a single dipolar interferometric point-spread function, where only the anisotropic part of the scattering contributes to the orientation-sensitive signal, while the isotropic part is intrinsically suppressed. This unique combination of unambiguous single-shot orientational readout, interferometric sensitivity, and anisotropy-selective contrast enabled us to achieve sub-degree angular localization precision and rotation tracking of low-drag ($\gamma_{rot} \sim 10^{-24}$ $J\ s$) orientation labels (DNA origami with attached gold nanorod) in an aqueous environment at microsecond temporal resolution. We demonstrate that cypiSCAT opens a measurement regime for direct torque detection down to ~1 pN nm, compatible with rotational tracking and torque measurement in single-molecule systems such as motor proteins.

# Results

## Interferometric PSF with cylindrically polarized scattered light

The cypiSCAT detection principle is based on a PSF-engineering approach, transducing the orientation of an anisotropic nanoscopic scatterer into the orientation of a dipolar interferometric PSF (Figure 1). To achieve this, we constructed a wide-field iSCAT microscope[31] incorporating a custom composite vortex half-wave plate, positioned in the plane conjugate to the image plane. The orientation of the half-wave-plate fast axis depends on the azimuthal coordinate $\theta$ as $\theta_{fast} = m\theta /2 + \theta_0$, where $m$ is a topological charge, and an optically isotropic segment positioned in the center, see Figure 1a. While the scattered wave is converted into a cylindrical vector beam by the vortex wave plate, the polarization of the reference wave transmitted through the central isotropic segment remains unchanged (Figure 1a-c).

The shape and contrast of the detected PSF are highly dependent on the degree of scattering anisotropy (Figure 1d, e, g). For a scatterer with an arbitrary degree of anisotropy, the scattered light can be expressed as a coherent sum of a purely anisotropic (Supplementary Note 1) and isotropic (Supplementary Note 2) contributions with amplitudes $s_{aniso}$ and $s_{iso}$, respectively. The intensity of the detected PSF can then be expressed as (see Supplementary Notes 1-3 for more details):

$$I_{PSF}(R,\theta) = I_{ref} + I_{int} + I_{scat} = r^2 + \sqrt{2} r s_{aniso} e^{-\frac{R^2}{w_0^2}} J_m(k_\perp R)[\sin 2\alpha(\cos m\theta \sin\varphi + \sin m\theta \cos\varphi) + \cos 2\alpha(\cos m\theta \cos\varphi - \sin m\theta \sin\varphi)] + \left(s_{iso}^2 + \frac{s_{aniso}^2}{2}\right) e^{-\frac{2R^2}{w_0^2}} J_m^2(k_\perp R). \quad (1)$$

Here, $r$ is the amplitude of the reference wave, $J_m$ is the Bessel function of the first kind and $m^{th}$ order (equal to the topological charge of the vortex retarder), $\alpha$ is the in-plane orientation of the anisotropic scatterer, and $\varphi$ is the interferometric phase.

The relative rotation of the interferometric term $I_{int}$, exhibiting a multipolar harmonic modulation along the azimuthal coordinate (Supplementary Figure 1), is dependent on the orientation of the anisotropic scatterer $\alpha$, interferometric phase $\varphi$, and topological charge $m$ *as* $\Delta\theta = (2\Delta\alpha + \Delta\varphi)/m$. The desired property of the detected PSF is the angular sensitivity of the PSF to the orientation of the anisotropic scatterer, which is highest for $m$=1 (used in this work), $(\Delta\theta/\Delta\alpha)_{m=1} = 2$, resulting in the dipolar interferometric PSF (Figure 1d,f).

The core mechanism behind the cypiSCAT PSF formation is the spatially varying interference of the cylindrically polarized scattered wave with the uniform circularly polarized reference wave (Figure 1b,c). The isotropic scattering yields scattered and reference waves having circular polarizations of opposite handedness. These orthogonally polarized waves do not interfere. The detected PSF is then a sum of a uniform background intensity $I_{ref}$ and the doughnut-shaped intensity of the cylindrically polarized scattered wave (Figure 1e), characteristic of cylindrical vector beams.[32] Anisotropic scattering, on the other hand, breaks this symmetry, enabling the interference. Due to cylindrical polarization, a certain part of the scattered beam always interferes constructively with the reference wave, while the opposite part, having opposite polarization, yields destructive interference. Combined, a dipolar PSF is formed (Figure 1d,f) with the orientation-sensitive interferometric term proportional to the anisotropic part of the scattering $s_{aniso}$, while isotropic scattering appears only in the latter intensity component, which diminishes rapidly at small scattering amplitudes, see Equation (1).

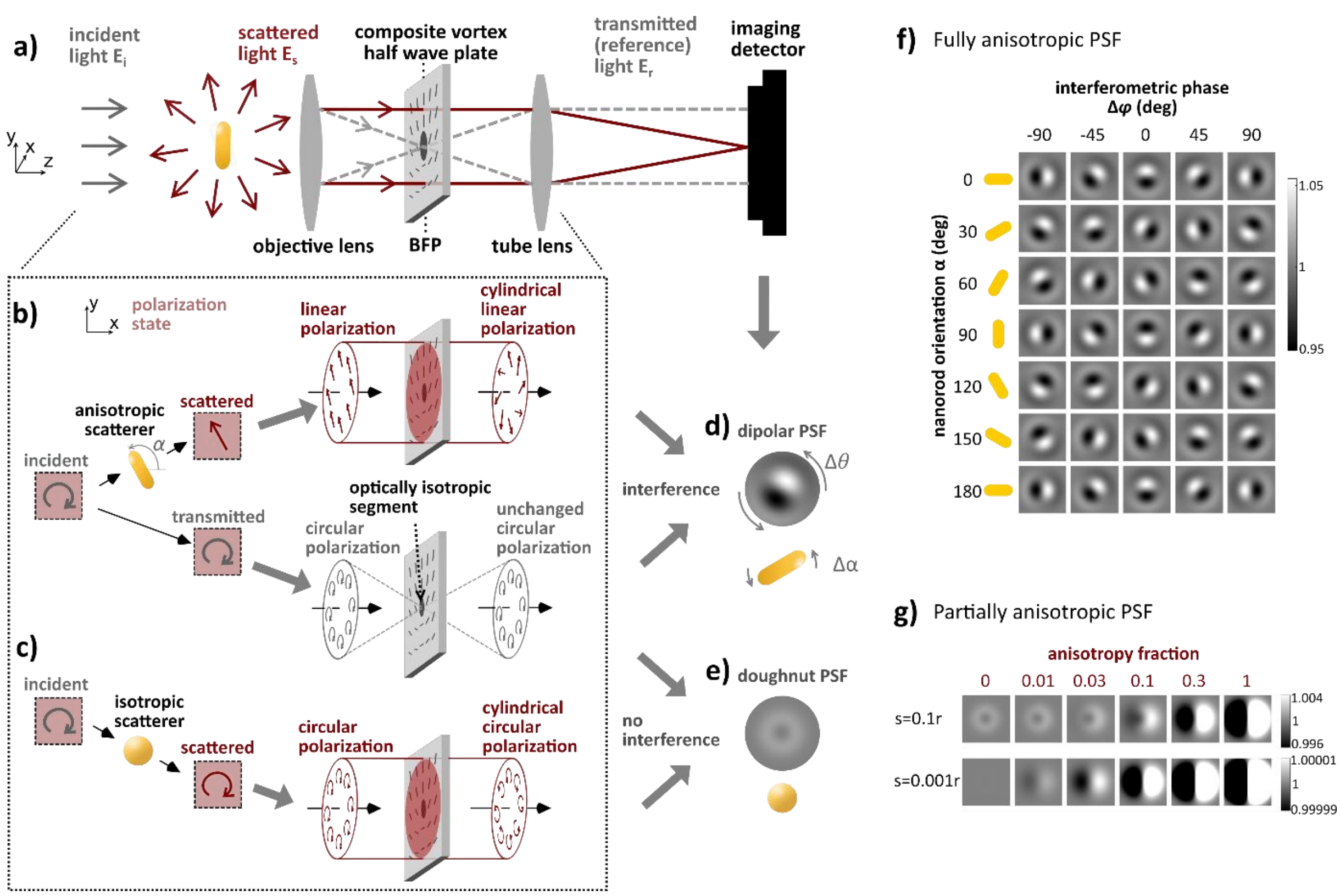

***Figure 1*** *Principle of the cylindrical-polarization-based iSCAT microscopy.* ***a)*** *Scheme of an iSCAT microscope with a composite vortex half-wave plate placed in its back focal plane (BFP). A light scattered* ***b)*** *anisotropically or* ***c)*** *isotropically by a subwavelength object is cylindrically polarized by the vortex WP, while the reference wave is transmitted through the central, optically isotropic, segment of the composite WP, resulting in either* ***d)*** *the dipolar interferometric PSF encoding the information about the scatterer orientation or* ***e)*** *a doughnut-shaped PSF.* ***f)****.Calculated PSFs of the anisotropic scatterer as a function of its in-plane orientation α and interferometric phase φ.* ***g)*** *Partially anisotropic PSFs calculated for different anisotropy fractions ($s_{aniso}$/( $s_{iso}$ + $s_{aniso}$)) of the scattered light (0 – fully isotropic, 1 – fully anisotropic scattering), and two cases of relative scattering strength, i.e., amplitude of the scattered wave at 10% and 0.1% of the reference wave.*

Assuming weak scattering (s << r), typical for deeply subwavelength scatterers such as single molecules or plasmonic nanoparticles, the scattering term in Equation (1), $I_{scat} \propto s^2$, becomes negligible compared to both the reference and interferometric terms, since $s^2 << rs << r^2$. Under these conditions, the dipolar, orientation-sensitive component of the PSF ($I_{int} \propto r\, s_{aniso}$) dominates over the doughnut-shaped contribution from isotropic scattering ($I_{scat} \propto s_{iso}^2 + s_{aniso}^2/2$), even at a low degree of scattering anisotropy, see Figure 1g. This decomposition highlights the operational advantage of cypiSCAT as it enables the orientation detection of nanoscopic objects even with subtle scattering anisotropy. In practice, the detectability of such scatterers is limited by the noise properties and polarization dependence (especially for high-numerical-aperture imaging) of the microscope setup.

## Composite vortex half-wave plate

To demonstrate this method experimentally, we have prepared a composite vortex half-wave plate by removing the central part of a liquid crystal polymer (LCP) layer from a commercial vortex half-wave plate (see Methods). The final element consists of the birefringent LCP layer with the orientation of the fast axis continuously changing along an azimuthal coordinate with the birefringence-free central disc, $r_{disc}$ = 0.75 mm, Figure 2a,b. The alignment of the central disc with the center of the vortex retarder was verified by the transmission patterns of the prepared composite element, placed between crossed or parallel polarizers, as shown in Figures 2c and 2d.

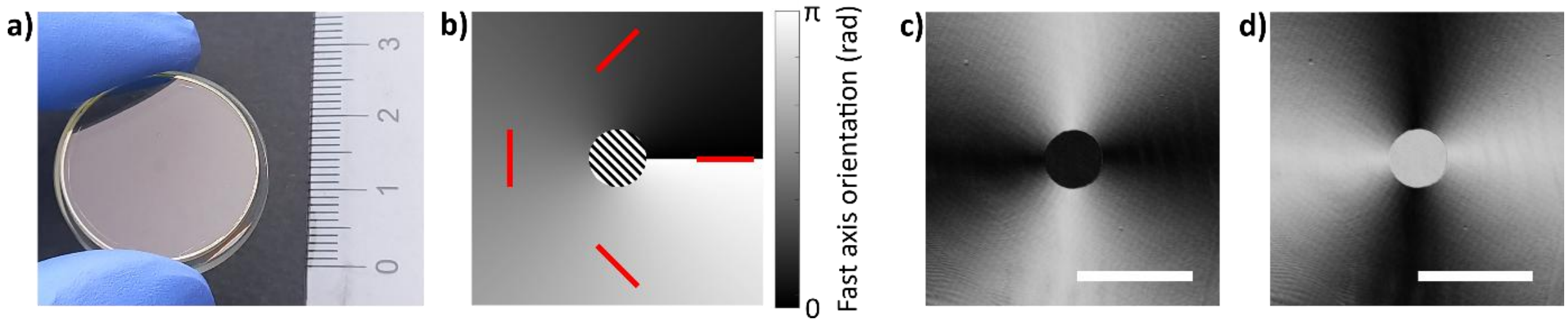

***Figure 2** Composite vortex half-wave plate. **a)** A photograph of the prepared composite polarization vortex element consisting of a birefringent polymer layer and etched circular hole sandwiched between glass substrates; **b)** The map of fast axis orientations (highlighted in red) of the LCP layer (not defined in the circular area without LCP). Measured transmission patterns of the polarization vortex element placed between two **c)** crossed and **d)** parallel polarizers. The scalebar length is 3 mm.*

## Single-frame orientation measurement of patterned nanorods

We tested the accuracy of the cypiSCAT method for determining the orientation of highly anisotropic scatterers using a set of gold nanorods patterned on a silicon wafer using electron beam lithography (see Methods). A scanning electron microscopy (SEM) image of four nanorods, oriented with 45-degree increments, is shown in Figure 3a. A representative cypiSCAT image, along with the corresponding fitted dipolar PSFs, is shown in Figure 3b. The absolute orientation of the nanorods can then be determined as $\alpha = (\theta_{PSF} + \theta_{const.})/2$, where $\theta_{const.}$ is dependent on the interferometric phase and the orientation of the composite vortex half-wave plate, see Supplementary Note 1. Assuming invariant $\theta_{const.}$ for each quartet of nanorods, the orientations of the nanorods were measured across multiple sets of patterned nanorods, see Figure 3d, yielding the determined rotation of the neighboring nanorods across the measured dataset to be (mean$\pm$SD): $\Delta\alpha = 45 \pm 6$ (deg), limited primarily by the fabrication precision of the nanorods.

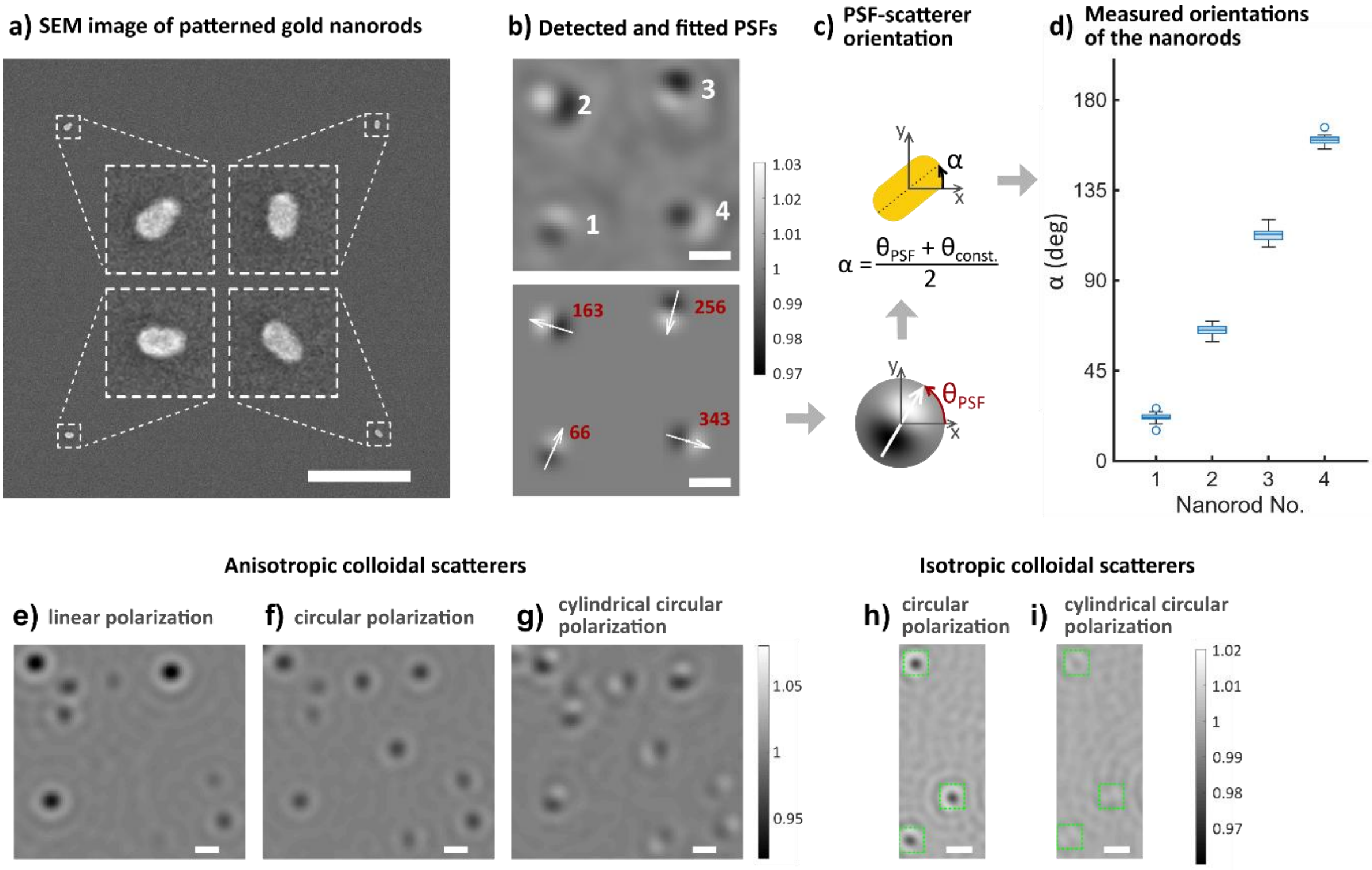

***Figure 3** Imaging patterned and colloidal nanoparticles with cypiSCAT. **a)** SEM image of the gold nanorods prepared using electron beam lithography (size ≈ 85 × 45 × 40 nm$^3$, length × width × height) on a silicon wafer. Orientation of the nanorods (rotated by 45 degrees by design) is highlighted in the zoomed images (insets). b**)** A representative cypiSCAT image with four dipolar PSFs corresponding to the quartet of rotated nanorods (top) and fitted PSFs with the determined orientations (bottom). **c)** A schematic representation of the relation between the PSF and nanorod orientations. **d)** Boxplot of the measured relative orientation of the imaged nanorods (obtained from N=18 images of 6 different nanorod quartets). **e)-g)** iSCAT images of colloidal gold nanorods (25 nm × 71 nm) immobilized on the glass coverslip acquired under different conditions - **e)** linear polarization, and **f)** circular polarization of the incident light (standard iSCAT, i.e., without the vortex retarder), and **g)** circular polarization of the incident light with vortex retarder (cypiSCAT). **h)** iSCAT images of colloidal gold nanospheres (60 nm diameter), acquired with circularly polarized illumination without the vortex retarder (standard iSCAT), and **i)** with the vortex retarder (cypiSCAT). Green boxes highlight the location of the nanospheres that become less visible in the cypiSCAT configuration. All scalebars correspond to 1 µm.*

Next, we have prepared a sample containing colloidal gold nanorods (GNRs, 25 nm × 71 nm) immobilized on the glass coverslip with random orientations. Interferometric images captured using standard iSCAT under linear and circular polarization of the illumination, as well as utilizing the cypiSCAT, are compared in Figure 3e-g. While the linearly polarized illumination yields the largest detected contrast for nanorods oriented close to the polarization vector (Figure 3e), it exhibits the highest variability in the detected contrast, as it is influenced by both the orientation and size of the nanorods. In standard iSCAT with circular illumination (Figure 3f), the information about the orientation is lost, and variations in the detected contrast stem from the size and shape heterogeneity, in line with the size distribution specified by the manufacturer. Finally, cypiSCAT (Figure 3g) provides disentangled information about the particle size and orientation, encoded in the PSF contrast and rotation, respectively.

The interferometric images of the isotropic scatterers were examined by detecting gold nanospheres (Figure 3h,i). As predicted above, see Equation (1), the elimination of the interferometric term for the isotropic part of the scattering amplitude suppresses the contrast of the isotropic scatterer in the cypiSCAT microscope (Figure 3h). Here, contrast of the gold nanospheres with 60 nm diameter detected by cypiSCAT (Figure 3h) is approximately 6-times lower than the contrast detected by standard iSCAT with circularly polarized illumination (Figure 3i).

## High-speed tracking of diffusively rotating tethered scattering labels

To demonstrate the power of the cypiSCAT method to track the rotational movements of single nanoscopic particles with high precision, we observed the diffusive rotation of a GNR tethered to a glass surface by a rotationally unconstrained linker (Figure 4a). To restrict the tip and tilt of the nanorod and keep the rotation axis closely aligned with the optical axis, we attached the GNR rigidly to a cross-shaped DNA origami[7] through multiple complementary DNA oligonucleotides (DNA-GNR, see Methods). The rapid diffusive rotation of tethered DNA-GNR orientation labels was then observed using a cypiSCAT microscope in transmission configuration (Supplementary Figure 9a), in which individual PSFs within a field of view were detected at 300,000 fps (Figure 4b). The detected frame-to-frame angular steps fall within the (-π/2, π/2) radians (Figures 4c,e), allowing for unambiguous unwrapping of the angular trace (see Supplementary Note 4).

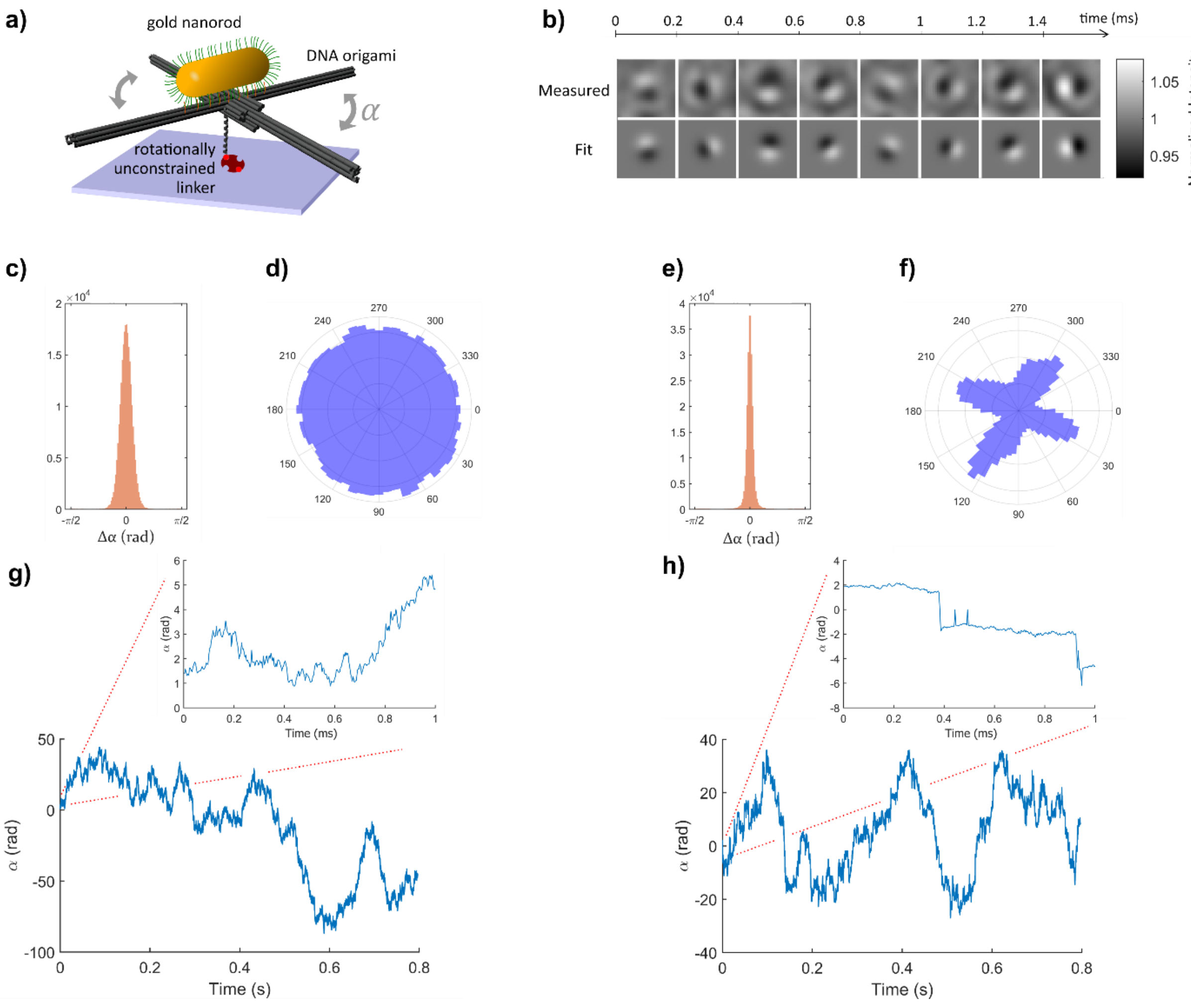

*Figure 4 Tracking the diffusive rotation of the DNA-GNR label. **a)** Schematic illustration of the experiment. A rotational label consisting of a gold nanorod attached to a cross-shaped DNA origami tethered to the glass surface through a rotationally unconstrained linker. **b)** Time series of the detected (top) and fitted (bottom) PSFs from single camera frames separated by 60 frames to*

*highlight the rotational dynamics.* ***c, e)*** *Distributions of the angular step sizes between consecutive frames, and* ***d, f)*** *polar histograms of the measured nanorod orientations, all taken from the whole trace (about 240,000 frames). Scalebar length is 200 nm.* ***g, h)*** *Representative angular traces of two distinct rotational labels in aqueous buffer, with the first 1 ms shown in insets.*

Measured angular traces of the DNA-GNR label (representative traces in Figure 4g,h) indicate a highly diffusive character of the rotational movement, driven by random thermal collisions. We observed 47 mobile orientation labels (0.8 s per trace) to determine one-dimensional rotational diffusion coefficients (see Methods). The statistical diversity of measured diffusivities is shown in Supplementary Figure 4. In Figures 4c-h, we show two representative examples highlighting unconstrained and partially constrained diffusive rotation. These are best evidenced by the differences between polar histograms (Figures 4d,f) and detailed angular traces in the inset of Figures 4g and 4h. The partially constrained rotation (Figures 4e,f,h) manifests as an increased residence probability of the label at specific angular positions, likely due to interactions between individual rotor blades and a feature on the surface. Overall, the label-to-label variability in the measured diffusion coefficients spans from 0 to 3000 $rad^2 s^{-1}$ (Supplementary Figure 4), with lower diffusivity values measured for the labels with increased interactions with the surface.

## Rotational microrheology from microsecond-resolved orientational fluctuations

Having access to orientational fluctuations of the nanoscopic orientation probe, we tested the reliability of the method for probing the local hydrodynamic parameters of the environment. We monitored rotational dynamics of single DNA-GNR label repeatedly in aqueous buffers with three different viscosities. Measured angular traces of a diffusively rotating label in an aqueous buffer with 0, 20, and 50% (v/v) glycerol are presented in Figure 5a in blue, orange, and green, respectively, together with traces of static gold nanorods immobilized on the glass surface (gray).

We compare two statistical approaches quantifying the diffusivity of the rotational motion (see Methods). The first is based on mean-square angular displacements (MSAD), plotted in Figure 5b, while the second considers angular step distributions (ASD), plotted in Figure 5c. The values of rotational diffusion coefficients ($D_{rot}$) obtained from both approaches, i.e., MSAD-derived (Figure 5d), and ASD-derived (Figure 5e), closely follow the theoretical dependence $D_{rot} \propto 1/\eta$ (Supplementary Note 6). However, the theoretical estimate of the diffusion coefficients based on the Tirado and Garcia de la Torre (TG) method[12] underestimates the experimentally measured $D_{rot}$ values (Supplementary Note 6), which we attribute to surface charge effects, as both DNA origami and DNA-functionalized gold nanorod are negatively charged. Another possible explanation, the plasmonic heating of the nanorod has been ruled out by simulations that showed negligible heating, $\Delta T \approx 0.2K$ under our experimental conditions (Supplementary Note 10).

The angular localization precision $\sigma_{loc}$ of the method can be estimated either from the standard deviation of the angular trace of the immobilized particle (gray lines in Figure 5a), or by extrapolating the MSAD plots to zero time lag, where (see Methods): $\sigma_{loc}^2 = (MSAD(0) + 2/3 D_{rot}\, t_E)/2$. The latter method accounts for the rotational motion blur of the PSF (see Figure 5f and Supplementary Note 8).

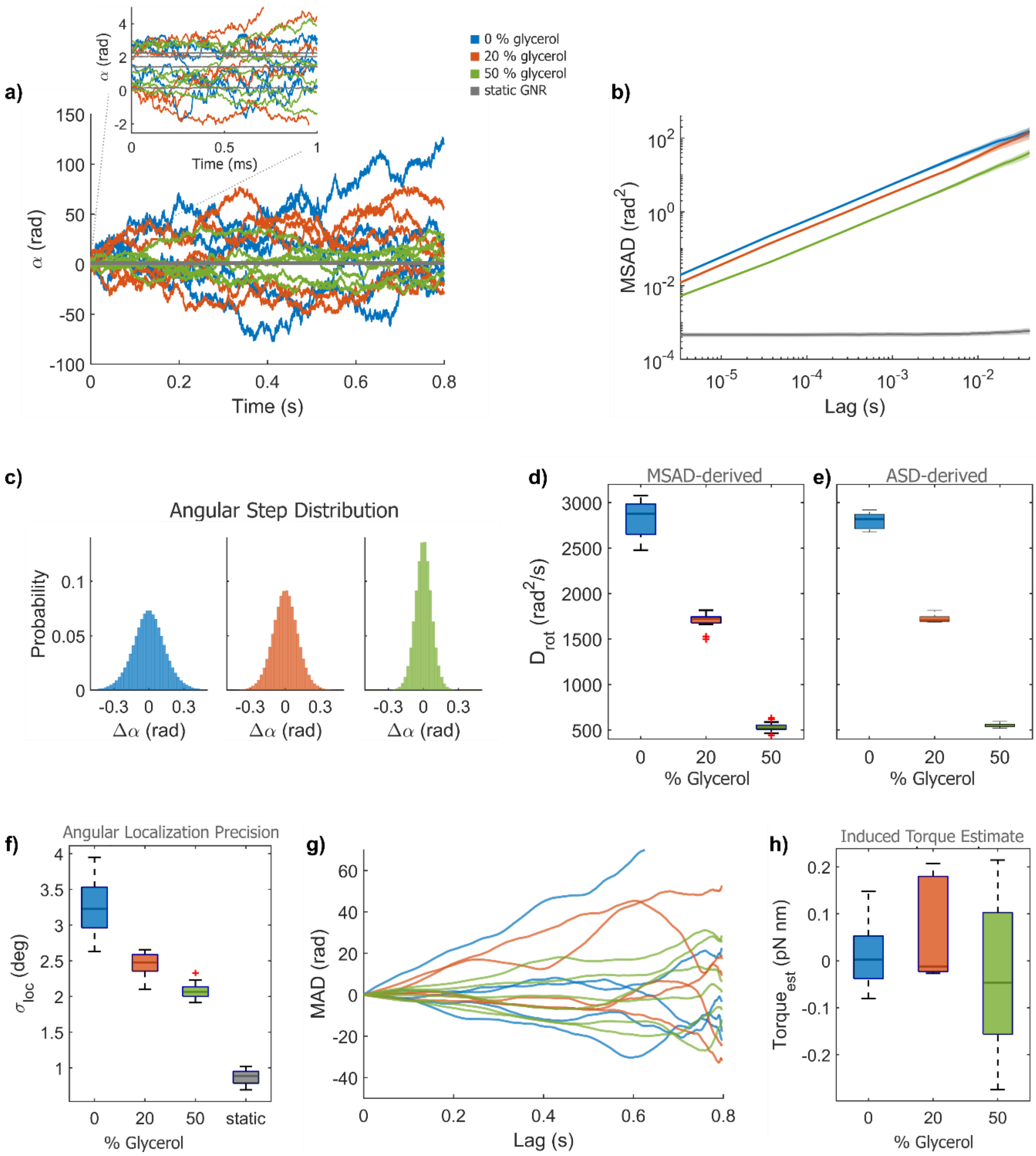

***Figure 5** Properties of diffusive rotation tracking at different buffer viscosities. **a)** Angular traces of the single rotational label measured at three different viscosities of the buffer (blue, orange, and green corresponding to 0, 20, and 50 %(v/v) of glycerol) and gold nanorod immobilized on the glass surface (gray lines). The first 1 ms of the traces are shown in the inset. **b)** Corresponding averaged mean square angular displacement plots with 95% confidence intervals (from $n_1$=15, $n_2$=15, $n_3$=18, $n_4$=12 angular traces) and **c)** angular step distributions. Rotational diffusion coefficients determined from **d)** MSAD plots and from **e)** angular step distributions. **f)** Angular localization precision determined at different rotational diffusivities of the DNA-GNR label. **g)** Mean angular displacement plots used to estimate the presence of torques driving the rotation of the label, and **h)** the boxplot of the estimated induced torques acting on the gold nanorod.*

## From angular traces to torque: calibration and sensitivity

Circularly polarized light can induce optical torque acting on asymmetric nanoparticles, such as gold nanorods[33], through the optical angular momentum transfer accompanying both absorption and scattering.[34] We use this interaction as a calibration means to characterize cypiSCAT as a quantitative torque-measurement platform. In the overdamped regime of the nanoparticle in liquid, an induced torque $\tau$ can be estimated from the angular velocity $\omega$ as $\tau = \omega\gamma_{rot}$, where $\gamma_{rot}$ is a hydrodynamic rotational drag. Here we first establish the baseline torque floor under standard imaging conditions and then quantify the dynamic range and calibration of optically driven torques as a function of applied intensity.

The angular velocities $\omega$ measured from the slopes of the mean angular displacement plots $(\mathrm{MAD} = \langle\alpha(t)\rangle = \omega t)$ in Figure 5g are rather randomly distributed around zero, indicating primarily thermally-driven rotations. Indeed, measured torque estimates at all three viscosities remain below $\tau$ = 0.3 pN nm, see Figure 5h, in line with the numerical estimate, $\tau_{opt,est}$ = 0.05 pN nm for our imaging conditions (Supplementary Note 11). Measured torques are far below the rotational equivalent of the energy of thermal fluctuations ($k_bT \approx$ 4.2 pN nm rad), confirming the dominant diffusive character of the observed rotational dynamics with minimum measurement-induced driving.

To calibrate the torque readout under controlled actuation, we introduce a second, circularly polarized laser beam ($\lambda$ = 640 nm, see Methods) focused onto a selected DNA-GNR label, without altering the imaging parameters ($\lambda$ = 660 nm, Figure 6a). Increasing the driving intensity produces a systematic bias in the angular traces (Figure 6b), while the stochastic thermal fluctuations remain apparent over millisecond timescales (Figure 6c). The transition from stochastic to torque-biased dynamics is quantified by the probability of measuring a biased angular velocity as a function of lag time in Figure 6d, indicating the timescale over which torque estimates become statistically reliable.

Combining the rotational tracking of multiple individual labels yields a calibration of the inferred optical torque (left axis in Figure 6e) and angular velocities (right axis) versus applied intensities of the driving beam (see also Supplementary Note 9). The measured torque increases approximately linearly with the driving intensity, with a fitted slope of $\tau$ = (1.2 ± 0.2) pN nm per 1 mW $\mu m^{-2}$ (Supplementary Figure 7c). The theoretical estimation suggests linear dependence of the optical torque of $\tau_{opt,est} \approx$ 2.0 pN nm per 1 mW $\mu m^{-2}$ (Supplementary Note 11), presented as a dashed line in Figure 6e. The residual spread in measured torques is consistent with the limited positioning precision of the focused Gaussian beam (spot size ≈ 320 nm) affecting the applied intensity of the driving beam.

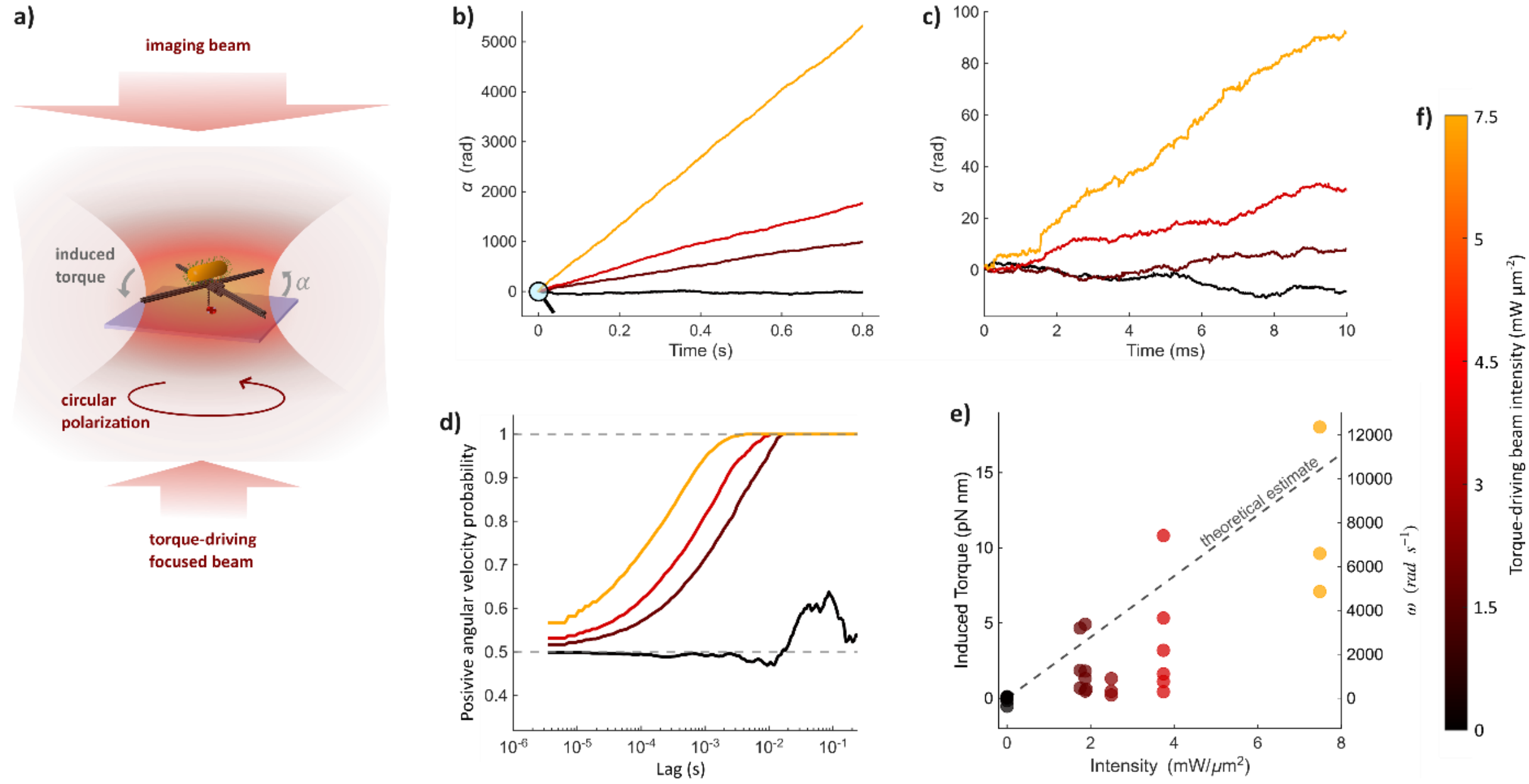

***Figure 6** Tracking rotation of gold nanorods spun by optically induced torque. **a**) Schematic illustration of the surface-attached DNA-GNR label rotated by the focused laser beam. **b)** complete 0.8 s-long and **c)** first 10 ms, highlighted by a magnifying glass in **b)**, of measured angular traces of an individual DNA-GNR label under different illumination intensities of the torque-driving focused beam. **d)** Plot of the measured angular velocity bias, i.e., probability that the measured angular velocity is biased (here positive) as a function of time lag. **e)** angular velocities (right y-axis) and optical torques (left y-axis) acting on the orientation labels measured at different intensities of the torque-driving focused beam. The theoretical estimate of the optical torque based on FDTD simulation is presented as a dashed line. **f)** A colorbar of the torque-driving beam intensities used in **b)**-**e)**.*

## Discussion

We introduce cypiSCAT, an interferometric scattering microscopy modality that uses cylindrical polarization to extract the orientation, position, and anisotropy of subwavelength particles from a single point-spread function. By directly linking the polarization of the scattered field to particle orientation, cypiSCAT provides precise and instant orientational readout. Furthermore, it is intrinsically selective to anisotropic scattering, manifested both theoretically (Figure 1g) and experimentally through the measured contrast drop of the isotropic nanosphere when switching from standard iSCAT to cypiSCAT (Figures 3h and 3i). This anisotropy-gated interferometric contrast provides built-in rejection of isotropic scattering backgrounds.

We tracked highly diffusive orientation labels driven by both thermal and optical torques, with angular localization precision down to two degrees (sub-degree precision for static nanorods). The microsecond-resolved (300,000 fps) angular traces yield drag and diffusion coefficients across different viscosities while explicitly accounting for angular localization noise and rotation blur. In this sense, cypiSCAT functions as a calibrated microrheology assay utilizing the rotational nanoscopic probe to report local hydrodynamic parameters. The high temporal bandwidth ensures a sampling density high enough for unambiguous unwrapping of angular traces. This enables reliable torque metrology demonstrated here by measuring optical torques down to 1 pN nm, establishing cypiSCAT as a powerful tool for probing nanoscale torque generation and dissipation in aqueous environments.

The ability to operate with compact, low-drag labels at microsecond temporal resolution is particularly relevant for observing rotational dynamics related to biomolecular machinery. While the F1-ATPase, generating torque of ~40 pN nm, has been extensively studied using a variety of orientation labels,[35] including micrometer-long actin filaments,[36,37] off-axis submicron beads,[38] gold nanoparticles,[39] or 500 nm Janus particles,[10] resolving individual reaction steps of DNA-processing motor proteins under physiological conditions remains elusive due to subtle torques involved (~10 pN nm for RNA polymerase[18,19]) and rapid reaction kinetics. Compared to recent high-speed scattering-based approaches, such as orientation analysis via dark-field microscopy with vortex wave plates[40] or polarization splitting[8], the interferometric nature of cypiSCAT enables significant miniaturization of orientation labels while retaining long observation times, improving access to weak torques and rare transient events.

Beyond engineered labels, the interferometric sensitivity of cypiSCAT extends to weak, non-plasmonic scatterers, for which even faint anisotropy preserves the dipolar, orientation-dependent PSF (Figure 1g). We anticipate that cypiSCAT will provide reliable orientation readout even when applied directly to anisotropic biomolecular complexes, suggesting a route toward label-free single-molecule orientation and localization microscopy.

# Methods

## Preparation of the composite vortex wave plate

A custom zero-order vortex half-wave plate (23 mm diameter, designed to operate at $\lambda = 660$ nm) prepared by continuous photoalignment of liquid crystal polymer (LCP) layer[41] was purchased from Thorlabs without the top cover glass. To introduce a segment without birefringent properties, the LCP layer was removed from the central area of the vortex retarder. First, we prepared a 1 mm thick cross-linked polydimethylsiloxane (PDMS) sheet (Sylgard 184) and cut a 1.5 mm hole into it. The PDMS sheet was placed on top of an LCP layer so that the center of the hole was aligned with the center of the vortex wave plate. The LCP layer was fully removed within the hole in the PDMS sheet using an oxygen plasma etching process in the Diener NANO plasma system (total 90 minutes, at 0.15 mbar process pressure and power set to 30%). The etched piece was placed on a cooled plate, and the temperature was monitored not to exceed 30°C during the process. The resulting vortex wave plate with the removed central part of the LCP layer was optically connected (optical adhesive NOA 61 from Norland Products, Inc.) with a cover window (2° round wedge window, PS810-A from Thorlabs) to protect the LCP layer from mechanical damage and deflect the parasitic reflections from the detection optical path.

## Preparation of the gold nanorods on silicon using electron beam lithography

Patterned arrays of gold nanorods were fabricated on silicon substrates by electron beam lithography (EBL). Silicon wafers were first cleaned and spin-coated with poly(methyl methacrylate) (PMMA, 3% in anisole) at 3000 rpm, followed by a soft bake at 180 °C for 90 s on a hotplate, yielding a resist thickness of approximately 120 nm. EBL patterning was performed using an electron lithography system (eLINEplus, Raith, Germany) operated at 20 kV with an exposure dose of 180 μC $cm^{-2}$. Samples were developed in a solution of methyl isobutyl ketone and isopropanol (MIBK:IPA, 1:3) for 40 s. Subsequently, a 40 nm thin gold layer was deposited via thermal evaporation in a vacuum (PLS 570, Pfeiffer Vacuum) at room temperature. Lift-off was achieved by sonication in acetone for 10 min at 50 °C, 80 kHz, and 50% power (Elmasonic P30H, Elma Schmidbauer GmbH, Germany), followed by rinsing with acetone and IPA, and drying under a stream of compressed nitrogen. The resulting nanorod arrays were characterized using scanning electron microscopy (Versa3D, FEI Company, USA).

## Optical setup of the cypiSCAT microscope

The optical layout of the cylindrical-polarization-based iSCAT microscope used in this work is based on wide-field illumination in either transmission or reflection geometry, see simplified schematics in Supplementary Figure 9. For experiments with surface-attached rotating orientation labels, transmission geometry was used (Supplementary Figure 9a) to minimize the effect of the interferometric phase on the orientation of the dipolar interferometric PSF, see Equation (1), and Figure 1f. A light emitted from a single-frequency, $\lambda$ = 660 nm, laser (Flamenco, Cobolt, Sweden) was coupled into a single-mode optical fiber (S405-XP), polarized circularly using a linear polarizer and a quarter-wave plate, and focused with a low NA on a liquid sample contained within a glass fluidic channel. Typically, illumination intensity within the field of view (7.3 × 7.8 $\mu m^2$) was 24 $\mu W\ \mu m^{-2}$. The light scattered by the subwavelength orientation labels was collected by the objective lens (NA = 1.3, UPLFLN100XOI2 from Olympus, Japan) and transmitted together with the incident beam (acting as a reference wave) towards the high-speed camera (iSPEED5, iX Cameras, UK) by means of series of mirrors (PF10-03, Thorlabs Inc., USA) and lenses in the detection optical path of the microscope. The focal lengths of the lenses were $f_1$=500 mm, $f_2$=400 mm, and $f_3$=300 mm (numbered from the objective lens to the camera), resulting in 208× microscope magnification. In experiments with immobilized colloidal nanoparticles, a different camera (MV1-D1024E, Photonfocus AG, Switzerland) was used, operating at about 1000 frames per second. The composite vortex wave plate was placed in the conjugate plane of the 4-f system formed by lenses L2 and L3.

The imaging of EBL-patterned gold nanorods on a silicon wafer was done in reflection geometry (Supplementary Figure 9b), where the illumination of the sample was done through the beam-splitter, and the light reflected from a silicon wafer was used as a reference wave.

In the experiments with induced optical torque, the optical setup based on transmission geometry (Supplementary Figure 9a) was modified by adding a λ = 640 nm laser (Rogue, Cobolt, Sweden) into the system. The mirror was replaced with a beam splitter, allowing for illumination of a selected spot within the imaged field of view, with the beam tightly focused by an objective lens. The illumination intensity was controlled by a gradient neutral density filter wheel, while the beam was introduced into the sample using a flippable beam stop just before starting the high-speed recording. A bandpass optical filter (FBH660-10, Thorlabs Inc., USA) was placed into the detection optical path to block the 640 nm beam from being detected by the camera.

## PEG passivation of glass surfaces

Glass coverslips (1.5H, thickness 170 ± 5 µm; Marienfeld, Germany) were immersed in methanol for 10 min, followed by incubation in a solution of 5 % (v/v) acetic acid and 1 % (v/v) aminosilane in methanol for 20 minutes. During this period, the coverslips were sonicated for 1 min (80 kHz, 50 W) in an ultrasonic bath. Coverslips were then rinsed with methanol, washed with Milli-Q water, and dried under a stream of compressed nitrogen. A solution of 3.125 mg/mL biotin-PEG-SC and 125 mg/mL m-PEG-SC (both MW = 5000 Da; Laysan Bio, USA) was prepared in freshly made 0.1 M sodium bicarbonate. A droplet of this solution was applied to two coverslips by bringing them into contact with the solution in the middle, with aminosilane-functionalized sides facing inward. The prepared coverslip sandwiches were kept incubating in the dark and humid environment overnight, then disassembled and rinsed with Milli-Q water before the iSCAT experiment.

## Assembly of the microfluidic channel

Microfluidic channels were fabricated using two glass coverslips, a bottom coverslip (22 mm × 22 mm, 1.5H, thickness 170 ± 5 µm from Marienfeld, Germany) and a top coverslip (18 mm × 18 mm). Both were first cleaned by sonication in isopropyl alcohol (10 minutes, 80kHz), then rinsed with ethanol, dried by N2, rinsed with MilliQ water, and dried again. The bottom coverslip was then treated with oxygen plasma for four minutes and passivated with PEG (see above) or left for subsequent BSA passivation (see below). Parafilm spacers (~100 µm thick) were placed between the coverslips, and the assembly was sealed by melting the parafilm on a hot plate at 100 °C for about 10 seconds, followed by gentle pressing of the top coverslip to form a tight seal.

## Preparation of the structured DNA rotors

The DNA origami rotors, ORBIT, originally designed for fluorescence-based rotational tracking of enzymatically unwound DNA,[7] were modified to enable attachment of a single gold nanorod on top of the cross-like structure, while having a rotationally unconstrained connection to a streptavidin-functionalized solid surface. The 100 nM structural DNA staple strands and 10 nM scaffold strand (M13mp18 viral DNA, single-stranded, New England Biolabs, USA), same as in the original publication[7], were mixed with 0.5-1 µM polyA nanorod-attachment strands (NR-1 – NR-6, see Supplementary Table 4) and 0.5-1 µM biotinylated surface-attachment strand in folding buffer (10 mM Tris, pH 8.0, 1 mM EDTA, and 18 mM $MgCl_2$). To fold the rotors, the oligo mixture was placed in a thermocycler and held at 80°C for 5 minutes, and then annealed by cooling, first to 65°C in 1°C steps every 5 minutes, then to 25°C in 1°C steps every 105 minutes. After annealing, the folded rotors were separated from the excess oligos using electrophoresis in a 2% agarose gel, and extracted from the gel using a Freeze 'N Squeeze spin column (BioRad).

## DNA functionalization of the gold nanorods

Gold nanorods (A12-25-650-CIT-DIH-1-25, Nanopartz Inc., USA) were functionalized with 5'-thiol-modified poly-deoxythymidine (5'-/5ThioMC6-D/$(T)_{30}$-3', HPLC-purified) from (Integrated DNA Technologies, Inc., USA) following a salt-aging and pH-modulation protocol, similar to previous reports[42]. Briefly, 10 µL of 100 µM polyT DNA oligo in PBS was mixed with 400 µL stock solution of gold nanorods (0.14 nM) in a 1.5 mL DNA LoBind Eppendorf tube (Eppendorf AG, Germany). First, salt concentration was increased by adding 4 µL of 1M NaCl (~10mM NaCl increase), followed by immediate vortex-mixing, and incubated for 60 minutes. Then, NaCl concentration was increased twice by ~120 mM by adding 58 µL and 66 µL of 1M NaCl in two consecutive steps, each followed by 60-minute incubation. Then, pH was changed from acidic to alkaline conditions by adding 20 µL of 1 M NaOH, resulting in a shift of pH from 4-5 to 8-9. The sample was diluted 1:3 in alkaline water (200 µM NaOH in DI water), centrifuged at 5000 × g for 5 minutes, and the supernatant was discarded. The pellet was resuspended in alkaline water to restore the original volume and thoroughly vortexed to recover GNRs adhering to tube walls. The centrifugation-resuspension step was repeated five to seven times to reduce unbound oligonucleotide concentration by several orders of magnitude. Functionalized GNRs were stored at 4 °C until further use.

## Preparation of the surfaces with diffusively rotating DNA origami-GNR labels

The microfluidic channel, with a PEG-passivated bottom surface, was mounted on an iSCAT microscope and flushed with 40 µL of TE buffer (10 mM Tris-HCl, 1 mM EDTA, pH 8.0). Next, 10 µL of 10 µM streptavidin in TE buffer was injected and incubated 10 minutes. This was followed by injection of 10 µL of 1nM DNA origami in TEM10 buffer (TE buffer with 10 mM $MgCl_2$), incubated for 20 min, and flushed with 100 µL of TEM10. Subsequently, 20 µL of 15 pM gold nanorods (GNRs) functionalized with single-stranded DNA oligonucleotides (5´-/5ThioMC6-D/ (T)30 -3´, HPLC purified) in TEM10 buffer was injected and incubated for 3 min. The channel was then flushed with TM1 buffer (10 mM Tris-HCl, 1 mM $MgCl_2$) to enable observation of free diffusive rotation of DNA origami-GNR labels. To modulate the viscosity, buffers containing 0-50 % (v/v) glycerol were introduced by exchanging the channel volume at least five times.

### Determination of the rotational diffusion coefficients from mean-square angular displacements

The values of $D_{rot}$ were calculated from $n_0$ = 47 angular traces of different DNA origami-gold nanorod labels (Supplementary Figure 4), and $n_1$=15, $n_2$=15, $n_3$=18 traces of the single label in aqueous buffer with 0 %, 20 %, and 50 % glycerol (Figure 5a). Unless stated otherwise, the diffusion coefficients were calculated from the corresponding MSAD plot at t = 330 µs time lag, i.e. 100 frames difference, indicated by gray dashed line in Supplementary Figure 5b. When the correction for the angular localization precision $\sigma_{loc}$, and the finite exposure time $t_E$ is accounted, the corrected $D_{rot}$ can be expressed as[43]: $D_{rot,corrected}(t) = (MSAD(t) - 2\sigma_{loc}^2)/(2t - 2/3\, t_E)$, see Supplementary Note 7.

### Determination of the rotational diffusion coefficients from angular step distributions

$D_{rot}$ coefficients can also be calculated from the angular step distributions (see Figures 5c and 5e). Here, considering again the angular localization precision characterized by $\sigma_{loc}$, and the finite exposure time $t_E$, the corrected rotational diffusion coefficient can be calculated as[43]: $D_{rot,ASD}(t) = (\sigma_{ASD}^2 - 2\sigma_{loc}^2)/(2t - 2/3\, t_E)$.

## Acknowledgements

We are grateful to Tomáš Špringer, who helped us with the functionalization of gold nanorods with DNA oligonucleotides, and Petra Lebrušková for thin film deposition.

This work was funded by the Czech Science Foundation, project No. 25-16414S, and the Ministry of Education, Youth and Sports, project ERC-CZ LL2409.

## Conflict of Interest

M.V., Ł.B., and M.P. have filed a patent application under the Patent Cooperation Treaty (PCT) related to the technique described in this manuscript.

## Author contributions

M.V., P.K., and M.P. conceived the project and secured funding; M.V. and V.L. developed the theoretical framework; L.T. and P.K. designed and prepared the DNA origami rotors; M.V., J.S., and V.L. performed the numerical simulations. L.B. and V.L. developed the acquisition software; M.V. processed the data. M.V. and I.K. performed the experiments; M.V. drafted the original manuscript, M.V., M.P., I.K., P.K., L.T., J.S. edited the manuscript and contributed to the final version. All authors contributed to the discussion of the results and interpretation of the data.

## Data availability

The datasets generated and analyzed in this study are stored on Figshare. A private access link is available from the first author upon reasonable request. Following acceptance in a peer-reviewed journal, the finalized dataset will be made publicly available on Figshare and assigned a DOI.

## Code availability

The MATLAB code used for image processing, PSF fitting, and statistical analysis of the processed data is stored on Figshare. A private access link is available from the first author upon reasonable request. Following acceptance in a peer-reviewed journal, the finalized dataset will be made publicly available on Figshare and assigned a DOI.

# Supplementary Information

## Supplementary Note 1: Dipolar interferometric PSF formation based on cylindrically polarized anisotropically scattered light

The orientation-sensitive detection principle presented in this work is based on transforming the anisotropically scattered light into a state with cylindrical polarization, interfering it with a reference wave having spatially uniform polarization, and detecting the resulting dipolar PSF (see Figure 1a). The incident light is circularly polarized to ensure equal illumination for all possible in-plane orientations of the anisotropic scatterers. Let's consider a left-handed circularly polarized incident light expressed using a Jones vector (similar results can be obtained for right-handed circular polarization):

$$\boldsymbol{E}_{inc}^{LHC} = \frac{1}{\sqrt{2}}\begin{pmatrix}1\\ i\end{pmatrix}. \qquad \text{(S1)}$$

As an example of a highly anisotropic scatterer, we will consider a plasmonic nanorod. The light scattered by the plasmonic nanorod near the resonance with the longitudinal surface plasmon mode is highly polarized along the longer direction of the nanorod. In a limit case of fully anisotropic scattering, the scatterer acts as a linear polarizer and the corresponding Jones vector of the scattered light is:

$$\boldsymbol{E}_{aniso}^{LHC} = \boldsymbol{M}_{aniso}\,\boldsymbol{E}_{inc}^{LHC} = s_{aniso}\begin{pmatrix}cos^2\alpha & cos\alpha\ sin\alpha\\ cos\alpha\ sin\alpha & sin^2\alpha\end{pmatrix}\frac{1}{\sqrt{2}}\begin{pmatrix}1\\ i\end{pmatrix} =$$

$$= \frac{s_{aniso}}{\sqrt{2}}\begin{pmatrix}cos^2\alpha + i\ sin\alpha\ cos\alpha\\ sin\alpha cos\alpha + i\ sin^2\alpha\end{pmatrix}. \qquad \text{(S2)}$$

Here, $\boldsymbol{M}_{aniso}$ is a scattering matrix of a fully anisotropic scatterer, $s_{aniso}$ denotes amplitude of the anisotropic scattering. We omit the plasmonic phase delay $\varphi_{SP}$ here (at resonance, ΔϕSP = π/2), as it will be included in the interferometric phase $\varphi$, as shown below. In the back-focal plane of the microscope, the scattered light is transmitted through a vortex half-wave retarder (vortex waveplate), which is a half-wave plate with the orientation of the fast axis $\theta_{fast}$ linearly dependent on the azimuthal coordinate $\theta$, i.e., $\theta_{fast} = m\theta/2 + \theta_0$, where $\theta_0$ is constant and $m$ is a topological charge of such vortex waveplate. A vortex waveplate of topological charge $m$, with the fast axis aligned with the $\boldsymbol{x}$ axis at $\theta = 0$ (i.e., $\theta_0 = 0$), see Figure 1, will transform the scattered beam into a cylindrical vector beam expressed as a function of azimuthal coordinate $\theta$:

$$\boldsymbol{E}_{aniso}^{CylPol}(\theta) = \boldsymbol{M}_{Vortex\,WP}\,\boldsymbol{E}_{scat}^{LHC} = \begin{pmatrix}\cos m\theta & \sin m\theta\\ \sin m\theta & -\cos m\theta\end{pmatrix}\frac{s_{aniso}}{\sqrt{2}}\begin{pmatrix}cos^2\alpha + i\ sin\alpha\ cos\alpha\\ sin\alpha cos\alpha + i\ sin^2\alpha\end{pmatrix} =$$

$$= \frac{s_{aniso}}{\sqrt{2}}\begin{pmatrix}\cos\alpha\,(\cos m\theta\cos\alpha + \sin m\theta\sin\alpha) + i\sin\alpha\,(\cos m\theta\cos\alpha + \sin m\theta\sin\alpha)\\ \cos\alpha\,(\sin m\theta\cos\alpha - \cos m\theta\sin\alpha) + i\sin\alpha\,(\sin m\theta\cos\alpha - \cos m\theta\sin\alpha)\end{pmatrix}. \qquad \text{(S3)}$$

The reference wave is not affected by the vortex waveplate as it is transmitted through a central part of the composite vortex waveplate, where the birefringent material is absent. The transmitted reference beam thus keeps its left-handed circular polarization. We will consider a phase-delay $\varphi$ between the scattered and reference wave on the detector, which in general includes the phase differences caused by different optical path lengths $\varphi_{OPL}$, plasmonic resonance-related phase change $\varphi_{SP}$ and the Gouy phase $\varphi_{Gouy}$, i.e., $\varphi = \varphi_{OPL} + \varphi_{SP} + \varphi_{Gouy}$. The reference wave is then:

$$\boldsymbol{E}_{ref}^{LHC} = \frac{re^{i\varphi}}{\sqrt{2}}\begin{pmatrix}1\\ i\end{pmatrix} = \frac{r}{\sqrt{2}}\begin{pmatrix}cos\varphi + i\ sin\varphi\\ -sin\varphi + icos\varphi\end{pmatrix}. \tag{S4}$$

Here, reference amplitude $r$ is introduced, $\boldsymbol{E}_{ref} = r\boldsymbol{E}_{inc}$, and the intensity of the reference wave is $I_{ref} = \boldsymbol{E}_{ref} \cdot \boldsymbol{E}_{ref}{}^{*} = r^2$. The cylindrically polarized scattered wave is imaged on the detector, interfering with the reference wave. The resulting interferometric PSF can be expressed as a complex dot product of the involved electric fields:

$I_{PSF} = \left(\boldsymbol{E}_{aniso}^{CylPol} + \boldsymbol{E}_{ref}^{LHC}\right) \cdot \left(\boldsymbol{E}_{aniso}^{CylPol} + \boldsymbol{E}_{ref}^{LHC}\right)^{*}$. Plugging equations (S3) and (S4) in, after simple algebraic manipulations we get:

$$I_{PSF,aniso}(\theta) = r^2 + \frac{s_{aniso}^2}{2} + \frac{rs_{aniso}}{\sqrt{2}}[sin2\alpha(\cos m\theta\ sin\varphi\ + \sin m\theta\ cos\varphi) + cos2\alpha(\cos m\theta\ cos\varphi - \sin m\theta\ sin\varphi)]. \tag{S5}$$

Here, $r^2 + s_{aniso}^2/2$ is a constant background intensity, and the intensity has $m$ pairs of interferometric maxima and minima positioned in the point-antisymmetric way around the PSF center, with their azimuthal orientation being a harmonic function of both nanorod orientation $\alpha$ and the interferometric phase $\varphi$.

So far, only azimuthal dependence has been considered. The dependence of the field of the cylindrical vector beam at its waist ($z$=0) can be approximated as[1]:

$$\boldsymbol{E}_{R}^{CylPol} = \boldsymbol{E}_0 e^{-\frac{R^2}{w_0^2}} J_m(k_{\perp}R) \tag{S6}$$

where $R$ is a radial coordinate, $w_0$ is the beam waist radius, $k_{\perp} = k\ sin\vartheta_0$, $\vartheta_0$ is the characteristic cone angle of the Bessel beam, $J_m$ is the Bessel function of the first kind and $m^{th}$ order (identical to the topological charge of the vortex waveplate defined above), $n$ is the refractive index, and $R$ is the radial coordinate. Taking into account both azimuthal (Eq. S3) and radial (Eq. S6) profiles, we can express the field of the cylindrically polarized scattered beam in focus as:

$$\boldsymbol{E}_{aniso}^{CylPol}(R,\theta) = \boldsymbol{E}_{R}^{CylPol}(R)\boldsymbol{E}_{aniso}^{CylPol}(\theta) \tag{S7}$$

And the intensity of the detected PSF is:

$$I_{PSF,aniso}(R,\theta) = r^2 + \frac{rs_{aniso}}{\sqrt{2}} e^{-\frac{R^2}{w_0^2}} J_m(k_{\perp}R)[sin2\alpha(\cos m\theta\ sin\varphi + \sin m\theta\ cos\varphi) + \\ + cos2\alpha(\cos m\theta\ cos\varphi - \sin m\theta\ sin\varphi)] + \frac{s_{aniso}^2}{2} e^{-\frac{2R^2}{w_0^2}} J_m^2(k_{\perp}R). \tag{S8}$$

The first term is the intensity of the reference wave, the second, interferometric term is a multipolar cylindrically harmonic function carrying the information about the orientation of the scatterer $\alpha$, interferometric phase $\varphi$, and topological charge $m$, see Supplementary Figure 1, and the third term is the doughnut-shaped intensity of the scattered wave.

Considering a weak scattering scenario, s << r, which is true for scatterers typically observed using iSCAT, such as single molecules or deeply subwavelength nanoparticles, we can neglect the pure scattering term as, $I_{scat} \propto \frac{s_{aniso}^2}{2} \ll \frac{r s_{aniso}}{\sqrt{2}} \ll r^2$. The shape of the PSF is then given by the interferometric term as the intensity of the reference wave is uniform.

In the case of $m$=1, interferometric PSF has a dipolar antisymmetric shape, with one intensity minimum and one intensity maximum positioned symmetrically to the PSF center. The rotation of this dipolar PSF is linearly dependent on both nanorod orientation $\alpha$ and the interferometric phase $\varphi$, see Figure 1f.

For a higher topological charge $m$ of the vortex half wave retarder, the PSF comprises of $m$ interferometric maxima and $m$ interferometric minima, while the sensitivity of the PSF rotation, $\Delta\theta$, caused by the changes of the orientation of the scatterer $\Delta\alpha$ as well as to the interferometric phase $\Delta\varphi$ is inversely proportional to $m$, i.e., $\Delta\theta = (2\Delta\alpha + \Delta\varphi)/m$, see Supplementary Figure 1 for m=1,2,3 and $\varphi = 0$.

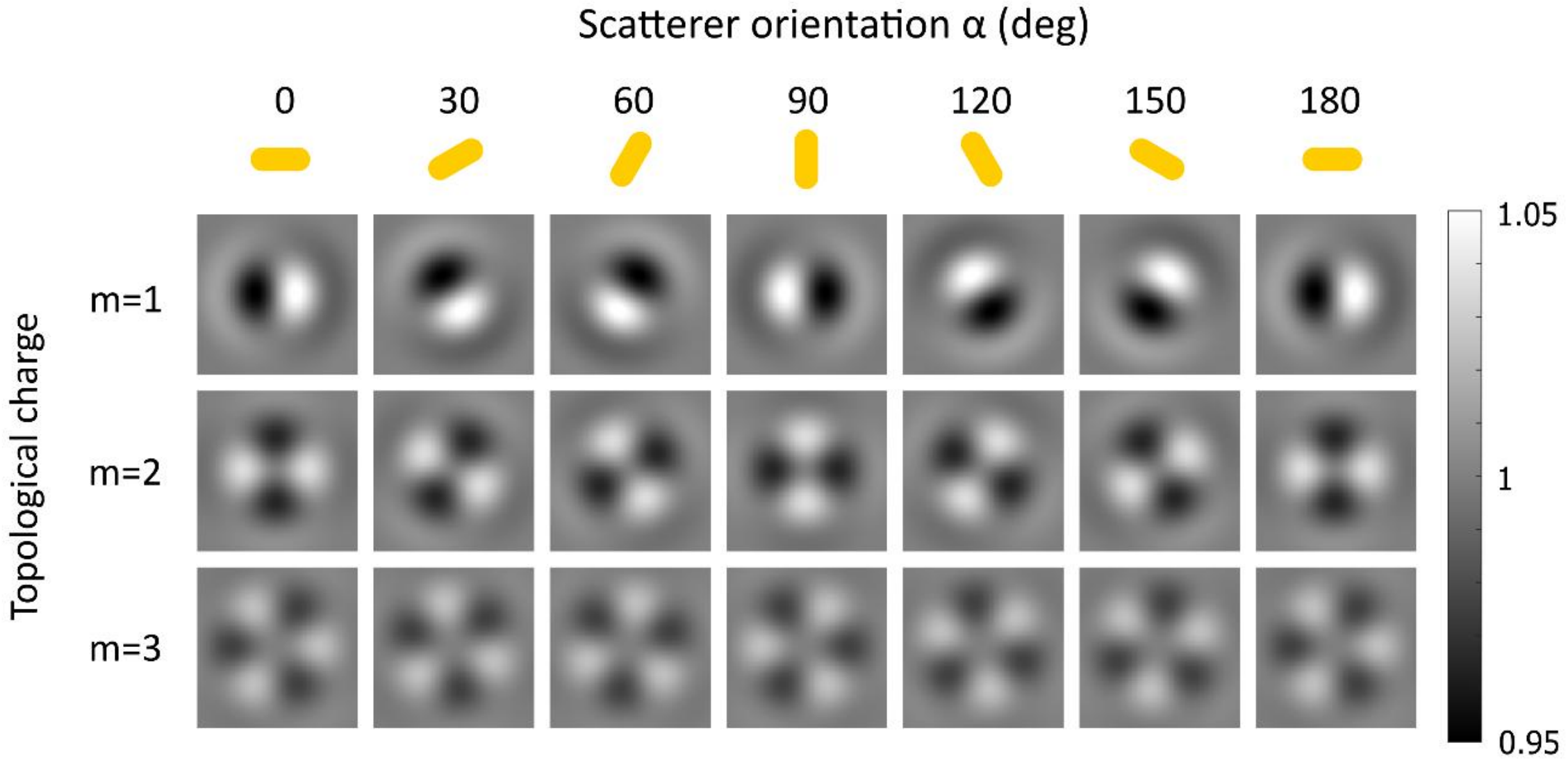

***Supplementary Figure 1*** *Dependence of the interferometric PSF shape and orientation on the anisotropic scatterer orientation for the first three values of topological charge of the vortex half wave retarder m=1,2,3. Amplitudes of the scattered and reference waves were* s *= 0.1, and r = 1, the interferometric phase* $\varphi = 0$.

## Supplementary Note 2: Doughnut shaped PSF formation based on cylindrically polarized isotropically scattered light

In case of purely isotropic scattering, the light is not linearly polarized by the scatterer as considered above, but rather keeps the polarization state of the incident light. In accordance with the anisotropic case above, we will again consider left-handed circularly polarized incident light.

$$\boldsymbol{E}_{iso}^{LHC} = \boldsymbol{M}_{iso}\, \boldsymbol{E}_{inc}^{LHC} = s_{iso} \begin{pmatrix} 1 & 0 \\ 0 & 1 \end{pmatrix} \frac{1}{\sqrt{2}} \begin{pmatrix} 1 \\ i \end{pmatrix} = \frac{s_{iso}}{\sqrt{2}} \begin{pmatrix} 1 \\ i \end{pmatrix}. \tag{S8}$$

Upon transmission through the vortex half wave retarder, its polarization state changes to cylindrical circular, while flipping the handedness of the circular polarization (see Figure 1c of the main text).

$$\boldsymbol{E}_{iso}^{CylPol} = \boldsymbol{M}_{Vortex\,WP}\, \boldsymbol{E}_{iso}^{LHC} = \begin{pmatrix} \cos m\theta & \sin m\theta \\ \sin m\theta & -\cos m\theta \end{pmatrix} \frac{s_{iso}}{\sqrt{2}} \begin{pmatrix} 1 \\ i \end{pmatrix} =$$

$$= \frac{s_{iso}}{\sqrt{2}} \begin{pmatrix} cos\, m\theta + isin\, m\theta \\ sin\, m\theta - icos\, m\theta \end{pmatrix}. \tag{S9}$$

Again, we can calculate the interferometric PSF as the intensity of the coherent sum of this cylindrically polarized wave (Eq. S9) with the reference wave (Eq. S4):

$$I_{PSF,iso} = \left(\boldsymbol{E}_{isoscat}^{CylPol} + \boldsymbol{E}_{ref}^{LHC}\right) \cdot \left(\boldsymbol{E}_{isoscat}^{CylPol} + \boldsymbol{E}_{ref}^{LHC}\right)^{*} = r^2 + s_{iso}^2. \tag{S10}$$

Here, the interferometric term is not present as the cylindrically polarized scattered wave and the reference wave have opposite handedness, i.e., have mutually orthogonal polarization states. Taking into account the radial component of the field (Eq. S6), we get a doughnut-shaped function:

$$I_{PSF,iso}(R,\theta) = r^2 + s_{iso}^2 e^{-\frac{2R^2}{w_0^2}} J_m^2(k_\perp R). \tag{S11}$$

## Supplementary Note 3: Hybrid asymmetric PSF of the partially anisotropically scattered light

In general, a light scattered by an arbitrary scatterer can be expressed as a linear combination of the fully anisotropic (Eq. S2) and fully isotropic (Eq. S8) contributions, resulting in the hybrid, partially asymmetric PSF as shown in the Figure 1g, and Supplementary Figure 2a.

This PSF of a scatterer with arbitrary degree of anisotropy can be expressed as:

$$I_{PSF}(R,\theta) = r^2 + \sqrt{2} r s_{aniso} e^{-\frac{R^2}{w_0^2}} J_m(k_\perp R)[\sin 2\alpha(\cos m\theta \sin\varphi + \sin m\theta \cos\varphi) +$$

$$+ \cos 2\alpha(\cos m\theta \cos\varphi - \sin m\theta \sin\varphi)] + \left(s_{iso}^2 + \frac{s_{aniso}^2}{2}\right) e^{-\frac{2R^2}{w_0^2}} J_m^2(k_\perp R). \tag{S12}$$

To assess the parametric dependence of the interferometric PSF shape on the scattering properties, in particular the scattering strength, the degree of anisotropy, and the dependencies on scatterer orientation α, and interferometric phase $\Delta\varphi$, we have written a Matlab script that facilitates the calculation and plotting of these PSFs.

As an example, the influence of the scattering strength (expressed as a ratio of the amplitudes of the scattered and reference waves *s/r*), and the degree of anisotropy (expressed as anisotropy fraction, i.e., 0 for a fully isotropic scattering and 1 for fully anisotropic scattering), is presented in Supplementary Figure 2.

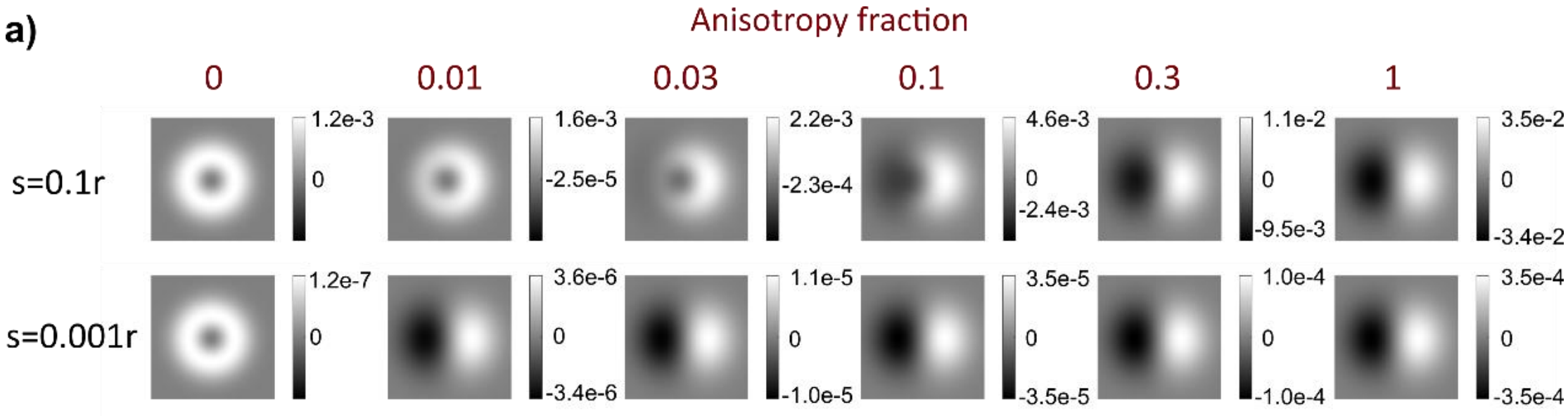

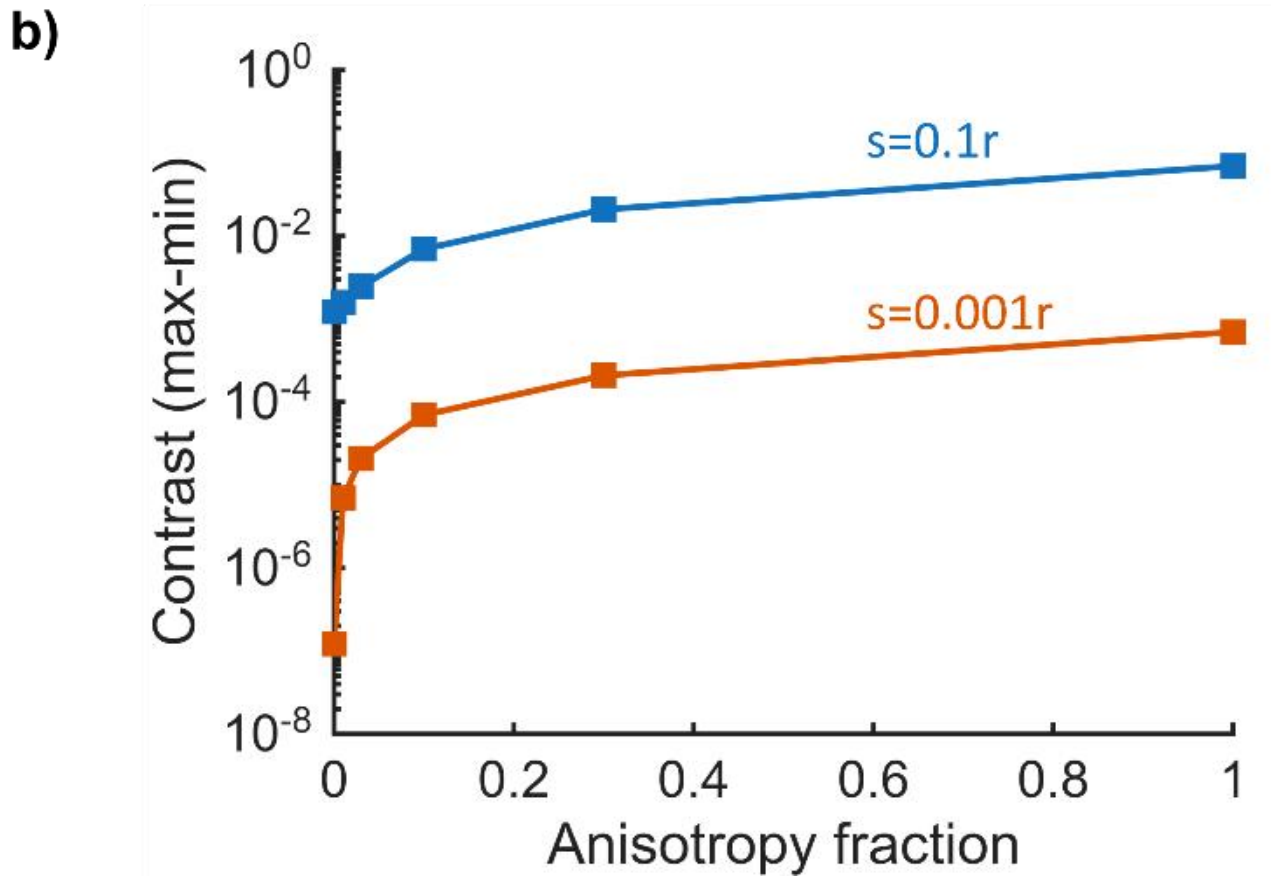

***Supplementary Figure 2 a)*** *Images and* ***b)*** *contrast of the interferometric PSF as a function of the scattering anisotropy fraction (0 = fully isotropic, 1 = fully anisotropic) calculated for two cases of the relative scattering strength, i.e., amplitude of the scattered wave at 10% and 0.1% of the amplitude of the reference wave, s = 0.1 r and s = 0.001 r, respectively.*

## Supplementary Note 4: Angular trace unwrapping

First, we determine the angular positions of the PSF by fitting the measured interferometric images with a dipolar function. The fitted angle $\theta_{PSF}$ is wrapped between 0 and 2π radians (Figure S1a) and needs to be unwrapped to determine the continuous angular trace. This can be done unambiguously when the size of the angular steps between two consecutive frames is not larger than π, i.e., $|\Delta\theta_{PSF}| < \pi$. Occasional jumps close to the ±2π are the artifacts arising from the PSF angle being wrapped between 0 and 2π, Figure S1b,c. Once the PSF angle is close to 0 or 2π, even small movement of the scatterer can cause the jump of the fitted angle over the 0-2π discontinuity. In our case, the 300 000 fps tracking speed was fast enough to observe even diffusive rotation of the origami-GNR label in the aqueous buffer (dynamic viscosity 0.93 MPa s), with the vast majority of the frame-to-frame angular steps being smaller than *π* radians. This enables us to unwrap the trace by recalculating the angles in the consecutive frames as follows:

$$\Delta\theta_{PSF} = \theta_{PSF}(n+1) - \theta_{PSF}(n) > \pi \quad \Rightarrow \quad \theta_{PSF}(n+1) = \theta_{PSF}(n+1) + 2\pi\,, \qquad (S12)$$

$$-\pi \le \Delta\theta_{PSF} = \theta_{PSF}(n+1) - \theta_{PSF}(n) \le \pi \quad \Rightarrow \quad \theta_{PSF}(n+1) = \theta_{PSF}(n+1) \qquad (S13)$$

$$\Delta\theta_{PSF} = \theta_{PSF}(n+1) - \theta_{PSF}(n) > \pi \quad \Rightarrow \quad \theta_{PSF}(n+1) = \theta_{PSF}(n+1) - 2\pi\,, \qquad (S14)$$

where *n* is the number of the frame.

Having the continuous angular trace of the PSF, we can use the relationship between the measured orientation of the interferometric dipolar PSF and the 2D orientation of the anisotropic scatterer, as described by Eq. (7). As we are interested in rotational changes rather than absolute orientation of the scatterer, we can neglect the orientation of the vortex half waveplate as well as interferometric phase and determine the angle of the anisotropic scatterer as $\alpha = \theta_{PSF,unwrapped}/2$. The unwrapped angular trace and the distribution of the frame-to-frame angular steps of the gold nanorod are presented in Supplementary Figure 1d and e.

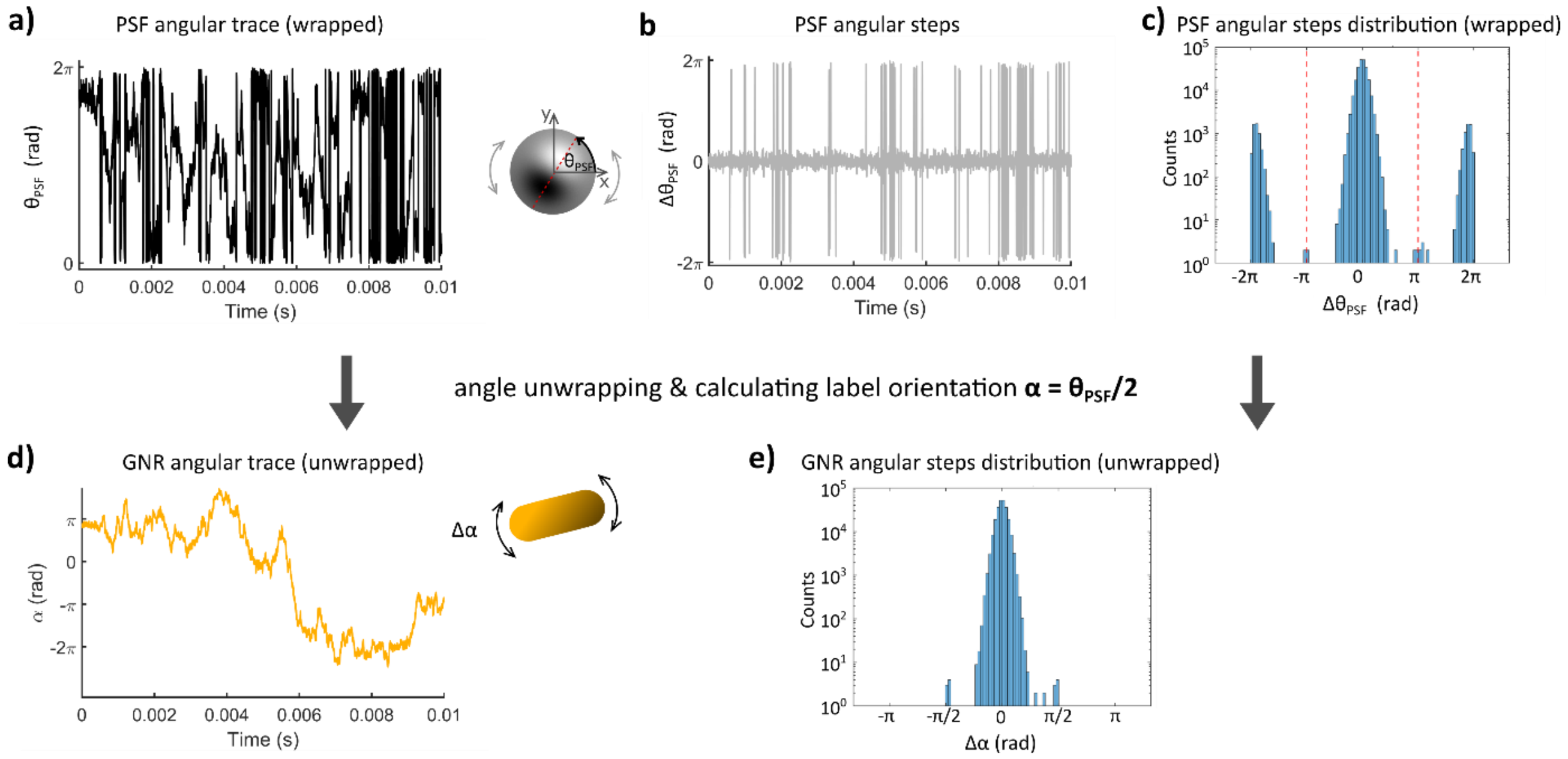

***Supplementary Figure 3*** *Process of calculating the label angular trace from the detected PSF orientations.* ***a)*** *Example of the angular trace of the fitted orientations of the PSF, wrapped between 0 and 2π radians.* ***b)*** *Time series of the angular steps between consecutive frames, and* ***c)*** *a histogram of the steps from the whole 240,000-frames-long wrapped PSF angular trace.* ***d)*** *Unwrapped continuous angular trace of the label, i.e., nanorod, and* ***e)*** *corresponding distribution of the angular steps between two consecutive frames.*

## Supplementary Note 5: Statistics of the measured rotational diffusion coefficients

The boxplot highlighting the diversity of the measured rotational diffusion coefficient $D_{rot}$ of the orientation labels in water is presented in Supplementary Figure 4. We have measured 47 traces, each about 0.8 s or 240,000 frames long. Although the dipolar PSF with high SNR was presented in the vast majority of detected frames, in a small fraction of frames (typically $<10^{-3}$, in best traces $<10^{-5}$), the PSF contrast dropped, probably due to the tip or tilt of the nanorod outside the sample plane, resulting in poor PSF fit. To avoid the determination of the diffusion coefficients to be biased by these non-reliable datapoints, each trace was segmented into shorter continuous subtraces consisting of frames where the PSF fit fulfilled specified thresholding parameters (PSF contrast, position within specified area, sum of residual squares below certain threshold). Mean square angular displacement (MSAD) plots were then constructed from these subtraces and $D_{rot}$ *coefficients* were calculated from the linear fits of these MSAD plots. Angular traces presented in the Figure 4 of the main text corresponds to the label number 23 (Figure 4c,d,e,i) and label number 16 (Figure 4f,g,h,j) from the Supplementary Figure 4.

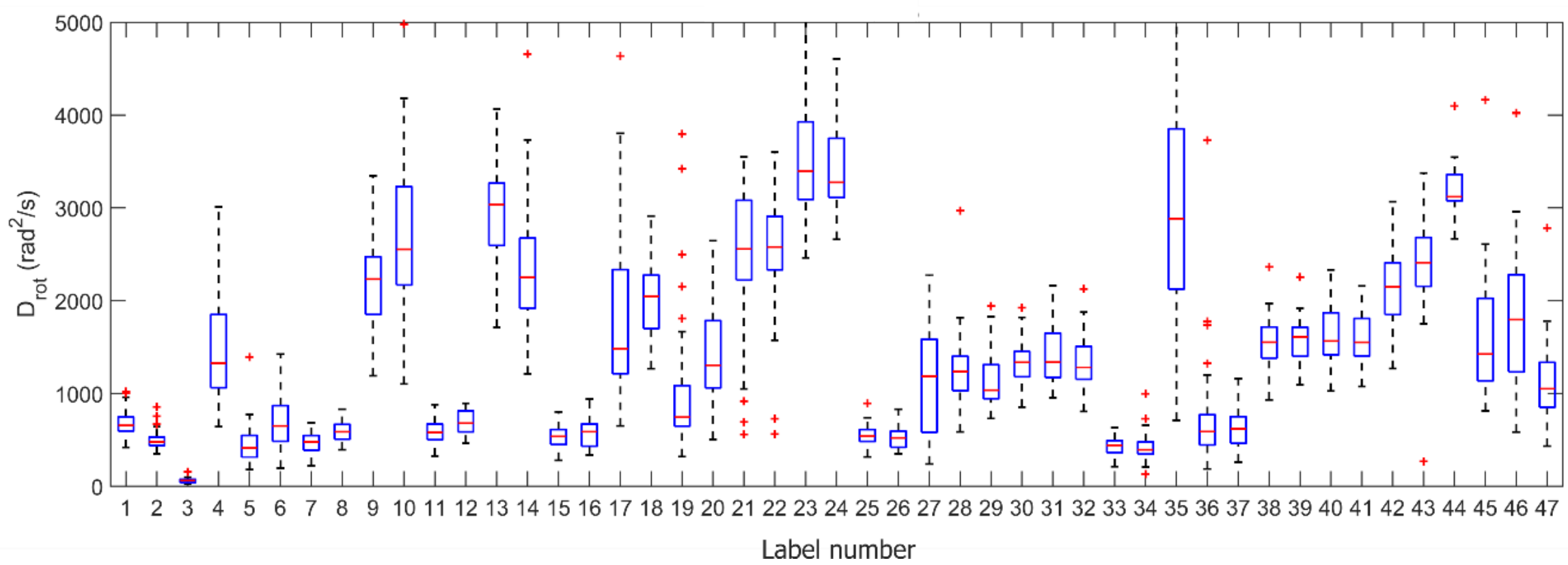

***Supplementary Figure 4*** *Statistics of the measured rotational diffusion coefficients from individual DNA origami-gold nanorod orientation labels (N=47) rotating diffusively around a rotationally unconstrained linker.*

## Supplementary Note 6: Estimation of the hydrodynamic drag and rotational diffusion coefficients

To estimate the total hydrodynamic drag of the orientation label used in this work, we assume both parts, i.e., gold nanorod and rotor-shaped DNA origami, can be approximated as a solid cylinder or a combination of solid cylinders. Assuming the 1D rotation of such a label around its center with the rotation axis perpendicular to both blades of the DNA origami rotor and gold nanorod, we can calculate the hydrodynamic rotational drag $\gamma_{rot}$ of each cylinder as[2]:

$$\gamma_{rot} = \frac{\pi \eta L^3}{3[ln(L/D)+\delta_\perp]}, \qquad \text{(S15)}$$

where $\eta$ is the viscosity of the liquid environment, $L$ and $D$ are the cylinder length and diameter, respectively, and $\delta_\perp$ is an end-effect correction dependent on the aspect ratio of the cylinder $L/D$. For a single DNA origami rotor blade, we assume $L$ = 160 nm, $D$ = 6 nm and the $\delta_\perp = -0.628$. For the nanorod, we assume $L$ = 71 nm, $D$ = 25 nm and the $\delta_\perp$ = -0.345, where the end-corrections were approximated as $\delta_\perp = -0.662 + 0.917(D/L) - 0.05(D/L)^2$.[3] Values of calculated hydrodynamic drag coefficients for the rotor-shaped DNA origami, 71 nm × 25 nm gold nanorod and the construct combining these two is summarized in Table S1 for three different dynamic viscosities of water-glycerol mixtures, i.e., $\eta_0 = 9.3 \times 10^{-4}$ $Nm^{-2}s$, $\eta_{20} = 18.2 \times 10^{-4}$ $Nm^{-2}s$, $\eta_{50} = 74 \times 10^{-4}$ $Nm^{-2}s$ *for* water with 0, 20, and 50 (v/v)% of glycerol, at room temperature ($T$ = 296 K).[4]

| **Supplementary Table 1** | **$\gamma_{rot}$ (N m s rad$^{-2}$)** | | |
|---|---|---|---|
| | GNR (71 x 25nm) | ORBIT DNA origami | ORBIT + GNR |
| water | 4.99E-25 | 3.00E-24 | 3.50E-24 |
| water:glycerol (4:1, vol) | 9.77E-25 | 5.88E-24 | 6.86E-24 |
| water:glycerol (1:1, vol) | 3.97E-24 | 2.39E-23 | 2.79E-23 |

From the Einstein-Smoluchowski relation, $D_{rot} = k_B T/\gamma$, we obtain an estimate of the rotational diffusion coefficients, as shown in Supplementary Table 2.

| Supplementary Table 2 | $D_{rot}$ (rad$^2$ s$^{-1}$) | | |
|---|---|---|---|
| | GNR (71 x 25nm) | ORBIT DNA origami | ORBIT + GNR |
| water | 8186 | 1360 | 1166 |
| water:glycerol (4:1, vol) | 4183 | 695 | 596 |
| water:glycerol (1:1, vol) | 1029 | 171 | 147 |

Above, the drag coefficients of the labels composed of subunits approximated as solid cylinders were calculated as a sum of drag coefficients of those subunits (Supplementary Eq. S15). However, this is only a rough approximation as the real drag would be smaller due to the hydrodynamic coupling. We have used HYDRO++ software[5] to simulate the rotational diffusion coefficients of complex shapes such as DNA cross-like structure with a gold nanorod more precisely. The model is based on approximating the shape of the construct using a set of spherical beads. Values of the $D_{rot}$ calculated using HYDRO++ are summarized in Supplementary Table 3.

| Supplementary Table 3 | $D_{rot}$ (rad$^2$ s$^{-1}$) | | |
|---|---|---|---|
| | GNR (71 x 25nm) | ORBIT DNA origami | ORBIT + GNR |
| water | 6074 | 1488 | 1354 |
| water:glycerol (4:1, vol) | 3104 | 760 | 692 |
| water:glycerol (1:1, vol) | 763 | 187 | 170 |

## Supplementary Note 7: Effect of the angular localization precision and finite exposure time on the measured rotational diffusion coefficients

To determine the rotational diffusion coefficients of the label from measured angular traces, we have applied two statistical-analytic approaches. The first is based on mean square angular displacement (MSAD), which in the case of the purely diffusive rotation along a single axis, is related to the rotational diffusion coefficient $D_{rot}$ analogously to the 1D translational diffusion, i.e., $MSAD = \langle(\alpha(t))^2\rangle = 2D_{rot}t$.[6] Here, $\alpha(t)$ denotes the difference between the angular positions measured at instants separated by time lag $t$. MSAD plots derived from the measured traces (Figure 5b, reproduced also here as Supplementary Figure 3a) show a linear dependence on the time lags over a range of four orders of magnitude. Thus, the calculated apparent $D_{rot}$ as a function of time lag is almost constant (solid lines in Supplementary Figure 3b). However, the measured angular positions and thus the values of mean square angular displacements are affected by the angular localization precision $\sigma_{loc}$ as well as the finite exposure time $t_E$. These effects can be accounted for to calculate the corrected rotational diffusion coefficient as[7]: $D_{rot,corrected}(t) = (MSAD(t) - 2\sigma_{loc}^2)/(2t - 2/3\, t_E)$, presented as dashed lines in the Supplementary Figure 3b. Comparing the apparent and corrected values of $D_{rot}$ in the Supplementary Figure 3b, we can see that the correction is significant only at the shortest lag times (note the logarithmic *x*-scale), where the effects of angular precision and motion blur within the exposure time of a single frame start to be comparable with the magnitude of the MSAD.

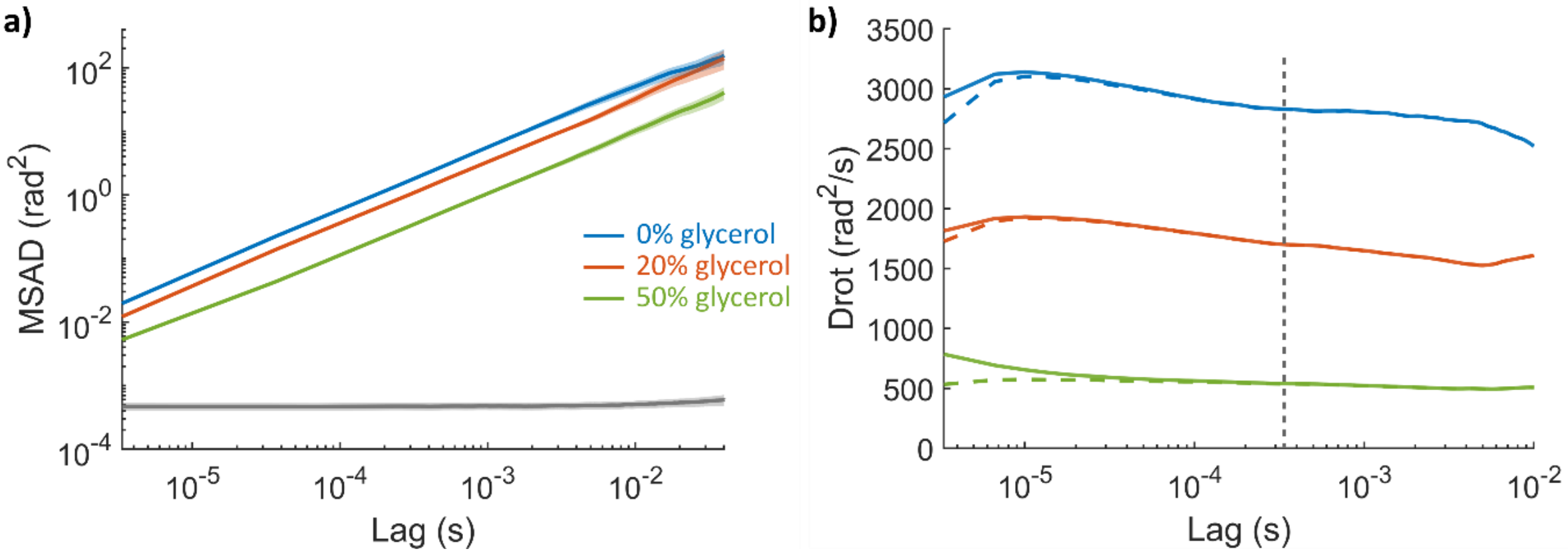

***Supplementary Figure 5*** *Calculation of the diffusional rotation coefficients from the MSAD plots.* ***a)*** *MSAD as a function of a time lag constructed from data measured with orientation labels at different viscosities of the buffer.* ***b)*** *Apparent (solid lines) and corrected (dashed lines) diffusional rotation coefficients calculated from the MSAD plots at different time lags. The vertical dashed line indicates a time lag of 330 µs (100-frames difference).*

Alternatively, $D_{rot}$ coefficients can be calculated from the angular step distribution (ASD, see Figure 5c and 5e). Here, considering again the angular localization precision characterized by $\sigma_{loc}^2$, the corrected can be calculated as $D_{rot,ASD}(t) = (\sigma_{ASD}^2 - 2\sigma_{loc}^2)/(2t - 2/3\, t_E)$.

## Supplementary Note 8: Effect of the rotational motion blur on the angular localization precision

The rotational mobility of the observed object may affect the precision of the angular position measurement due to rotational motion blur. This effect becomes substantial when tracking low-drag labels in liquids presented in this work. Indeed, we observed a dependence of the angular localization precision on the diffusivity in our experiments even at 300,000-fps framerate, see Figure 5f. We have achieved sub-degree angular localization precision, $\sigma_{loc}$ = (0.9 ± 0.1) deg, of the static gold nanorod, while the orientation of the diffusively rotating nanorods was measured with $\sigma_{loc}$ = 2.1 ±0.1 deg , 2.5 ± 0.2 deg, and 3.2 ± 0.4 deg at increasing rotational diffusivities, $D_{rot}$ of about 540, 1700, and 2830 $rad^2\ s^{-1}$ (calculated from mean MSAD-derived values, see Figure 5d). This diffusivity-related angular localization precision can be attributed to the rotational motion blur, which causes a shape change and contrast drop of the PSFs detected within the exposure time of each acquired frame, in our case, $t_E$ = 2.7 µs. The mean angular displacement of the PSF $\langle \Delta\theta_{blur} \rangle$ causing the motion blur can be estimated from the mean angular displacement of the diffusively rotating object $\langle \Delta\alpha_{blur} \rangle$ in our case (1D rotation): $\langle \Delta\theta_{blur} \rangle = \langle 2\Delta\alpha_{blur} \rangle = \sqrt{8 D_{rot} t_E}$. For the abovementioned measured diffusivities of the label, the mean rotational motion blur is then $\langle \theta_{blur} \rangle$ = 14 deg, 11 deg, and 6.2 deg.

## Supplementary Note 9: Measuring the rotational dynamics of DNA origami-gold nanorod labels driven by optical torque

The information-rich angular traces (Supplementary Figure 6b,c) of the DNA origami-gold nanorod labels (Supplementary Figure 6a) allows us to determine the distribution of apparent angular velocities for time lags from microseconds to seconds (Supplementary Figure 6d,e). Apparently, while the mean values of the measured angular velocities are almost constant over five orders of magnitude of the measurement time (Supplementary Figure 6f), the standard deviation of the angular velocities drops fast with time (Supplementary Figure 6g).

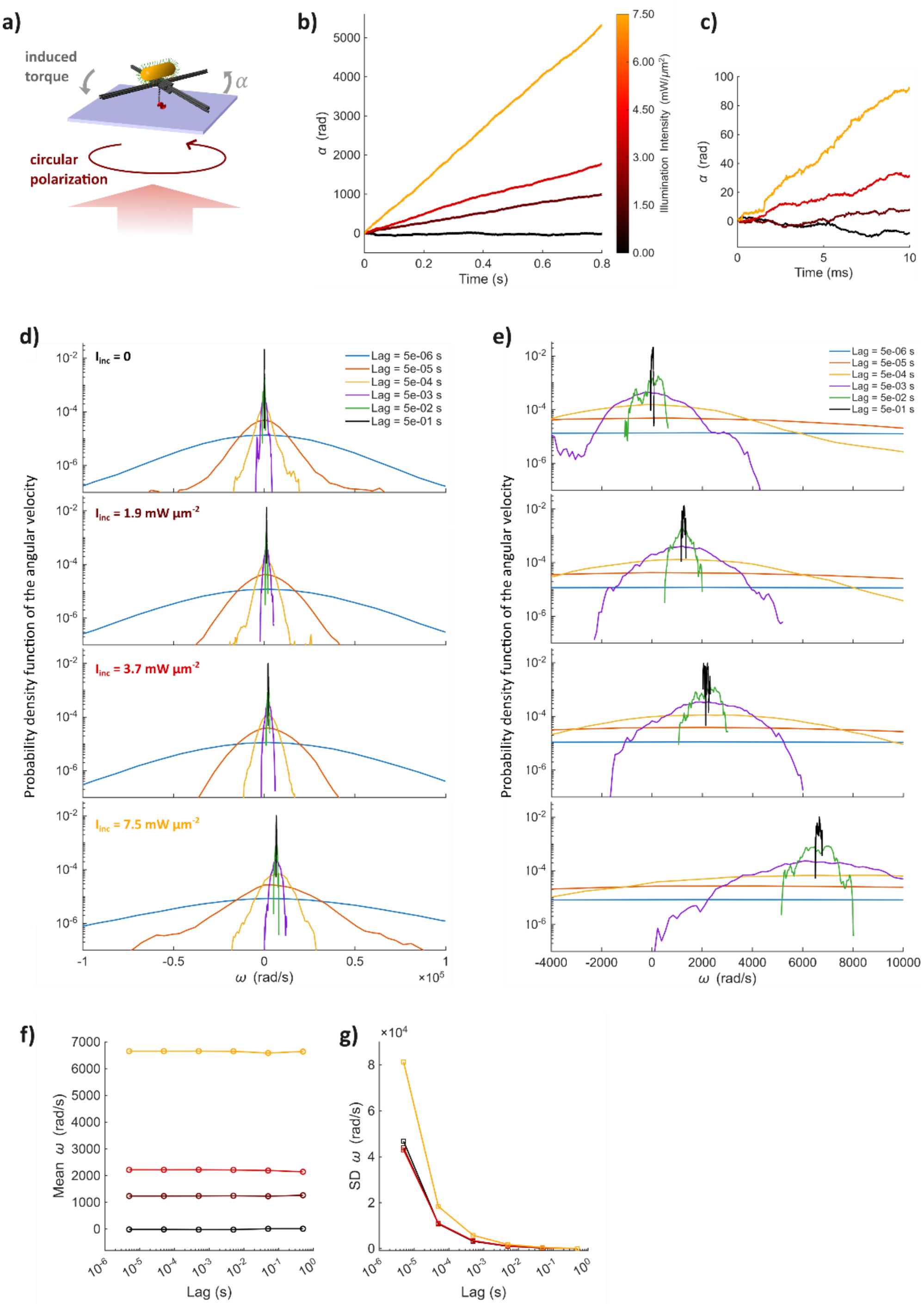

***Supplementary Figure 6*** *Tracking rotation of gold nanorods spun by optically induced torque.* ***a)*** *Schematic illustration of the surface-attached DNA origami label with a gold nanorod rotated by the focused laser beam.* ***b)*** *Whole, and* ***c)*** *first 10 ms of measured angular traces of an individual DNA origami-GNR label under different illumination intensities of the torque-inducing beam.* ***d)*** *Distributions of the measured angular velocities for time lags from 5 µs to 500 ms plotted over a broad range of velocities ($-10^5$ to $10^5$ rad $s^{-1}$) and* ***e)*** *over a smaller interval, highlighting the shift of the distributions towards higher mean values for increasing intensity of the torque-driving, λ=640 nm, circularly polarized illumination.* ***f)*** *Corresponding means and* ***g)*** *standard deviations of the angular velocities as a function of time lag.*

The complete dataset for the optical torque-induced rotational dynamics, i.e., 28 traces recorded at different intensities of the λ=640 nm driving beam and multiple (N=10) individual orientation labels, is presented in Supplementary Figure 7. Here, measured angular traces (Supplementary Figure 7a) and corresponding mean angular displacement plots with linear fits (Supplementary Figure 7b) were used to determine the angular velocities (right vertical axis of the Supplementary Figure 7) and optical torques (left vertical axis).

Our experimental conditions allowed us to observe the induced rotation of DNA origami-gold nanorod-based orientation labels only up to certain threshold intensities. In some experiments with $I_{inc,640}$ > 3 mW µm$^{-2}$ and all experiments with $I_{inc,640}$ > 7.5 mW µm$^{-2}$, we observed either stalled rotation or label unbinding from the surface, indicating heat-related unfolding of the DNA origami structure. This observation is consistent with the numerical simulations of plasmonic heating of gold nanorod in water with temperature increases of about $\Delta T$=70 K calculated for $I_{inc,640}$=7.5 mW µm$^{-2}$ mean illumination intensity (corresponding to 15 mW µm$^{-2}$ peak intensity of the focused gaussian beam, see Supplementary Figure 8).

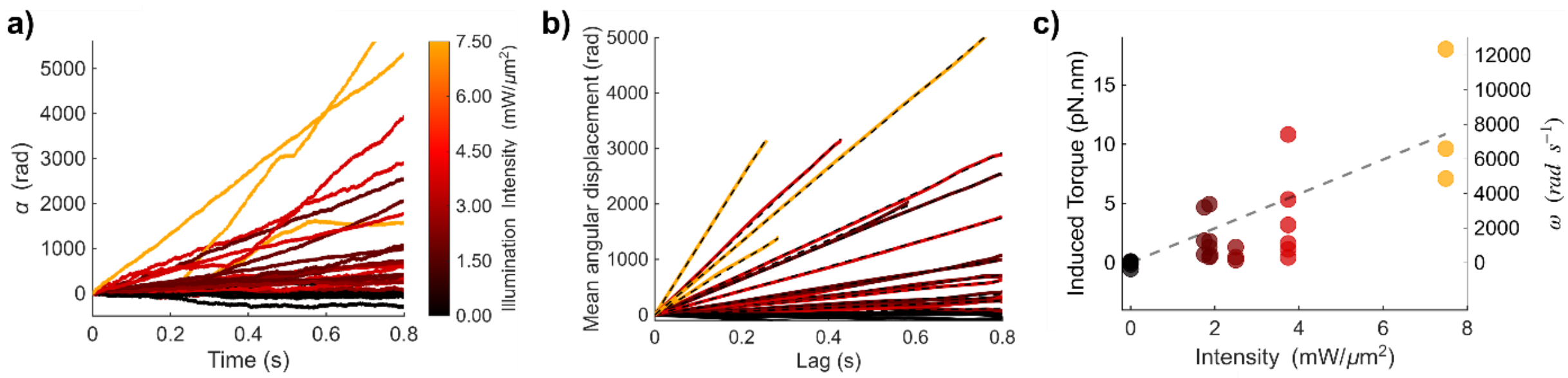

***Supplementary Figure 7 a)*** *All measured angular traces (N=28) recorded under different intensities of the torque-inducing (λ=640 nm) laser beam.* ***b)*** *Corresponding mean angular displacement plots with linear fits (dashed lines) used to determine the optical torques.* ***c)*** *Measured torques (left vertical axis) and mean angular velocities (right vertical axis) calculated from linear fits of the mean angular displacement plots.*

## Supplementary Note 10: Estimation of the plasmonic heating during the rotational tracking experiments

To estimate the effect of plasmonic heating on the diffusive motion of gold nanorods, we have numerically simulated the plasmonic heating of a single gold nanorod in water using COMSOL.

The nanorod is modeled as a 25-nm diameter cylinder with hemispherical caps, with a total length of 71 nm. The simulation volume (water) is a cylinder 40 times bigger than the dimension of the nanoparticle. The initial temperature of the system is set at $T_0$ = 296 K. The spectrum of absorption cross-section $\sigma_{abs}$ of a 25 × 71 nm gold nanorod, obtained using an FDTD simulation (Ansys Lumerical FDTD), is depicted in Supplementary Figure 6a. We have simulated two cases of plasmonic heating with parameters used in the experiments. The first, low-intensity imaging illumination, $I$ = 24 µW/µm$^2$ at $\lambda$ = 660 nm, $\sigma_{ab}$ = 1740 nm$^2$, Supplementary Figure 6b, and the second, optical torque inducing illumination, $I$ = 15 mW/µm$^2$ at $\lambda$ = 640 nm, $\sigma_{abs}$ = 1000 nm$^2$, Supplementary Figure 6c, d. The absorbed power is applied volumetrically within the gold nanorod as a uniform heat source, and the steady-state heat transfer problem is considered, based on the previously reported modelling approach.[8]

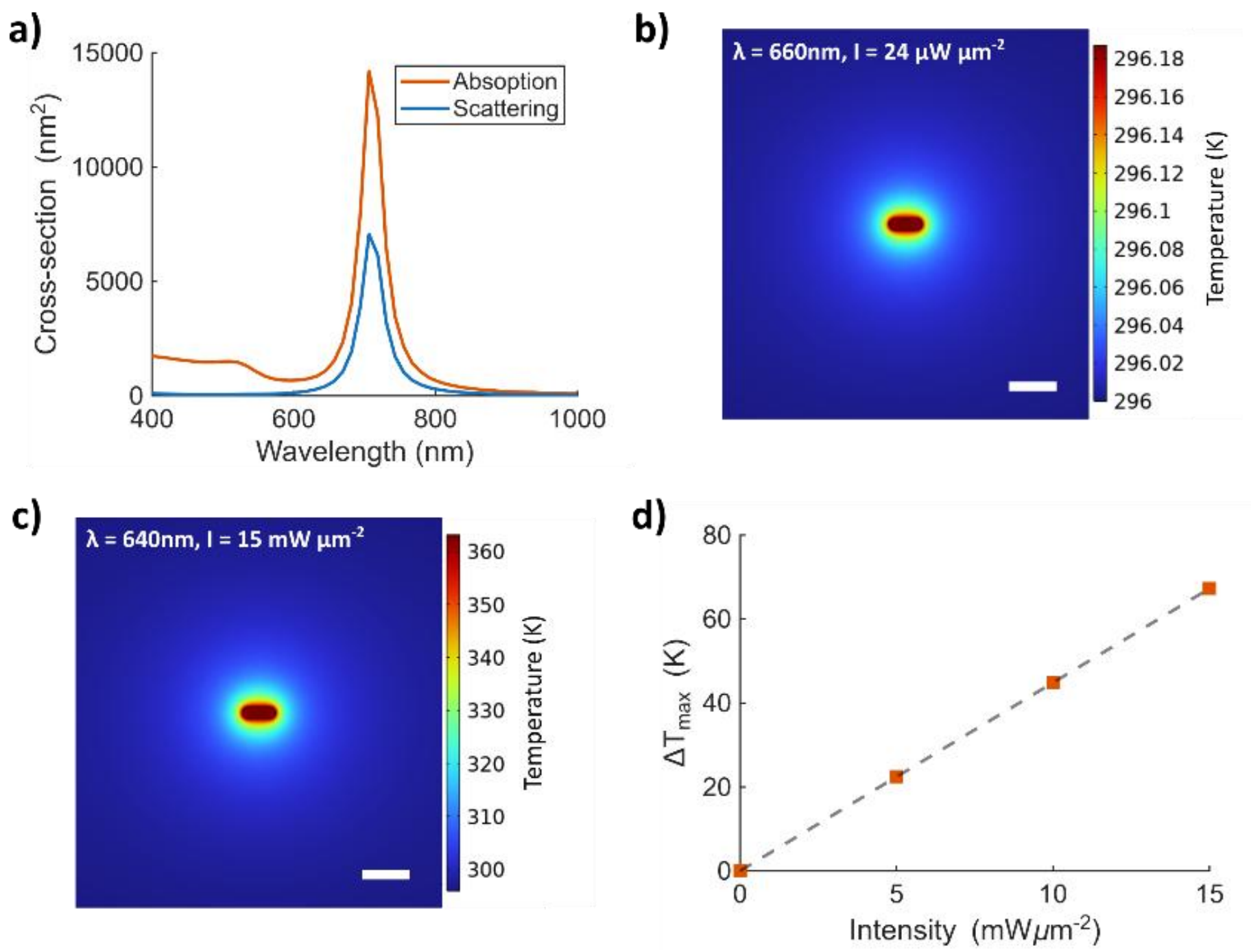

**Supplementary Figure 8** Numerical simulations of the gold nanorod plasmonic heating. **a)** Wavelength dependence of the scattering and absorption cross sections of the 25 nm × 71 nm gold nanorod in water, numerically calculated for a circularly polarized illumination. Steady-state temperature distribution of a gold nanorod in water, illuminated with circularly polarized light, **b)** I=24 μW μm$^{-2}$, λ=660 nm, and **c)** I=15 mW μm$^{-2}$, λ=640 nm. **d)** Maximum temperature change of the gold nanorod induced by circularly polarized light**,** λ=640 nm, as a function of the incident intensity. The scale bars in b) and c) correspond to 100 nm.

## Supplementary Note 11: Simulation of the optical torque acting on a gold nanorod

The optical torque exerted on the plasmonic nanorod was theoretically estimated using the Maxwell Stress Tensor (MST) formalism.[9]

The time-averaged optical torque $\boldsymbol{\tau}$ induced by incident circularly polarized light was computed by integrating the cross product of the position vector $\boldsymbol{r}$ and the electromagnetic surface force density ($\boldsymbol{T}\cdot\boldsymbol{n}$) over a closed surface S enclosing the particle: $\boldsymbol{\tau} = \oiint_S \frac{1}{2} Re(\boldsymbol{r} \times \boldsymbol{T}.\boldsymbol{n}) dS$. Here, $\boldsymbol{T}$ denotes the Maxwell stress tensor, defined as $T_{ij} = \varepsilon E_i E_j^* + \mu H_i H_j^* - {}^1/_2\, \delta_{ij}(\varepsilon|\boldsymbol{E}|^2 + \mu|\boldsymbol{H}|^2)$, where $E_{i,j}$ and $H_{i,j}$ are electric and magnetic field components ($i$, $j$ = x, y, z), $\varepsilon$ and $\mu$ are permittivity and permeability of the surrounding medium, and $\boldsymbol{n}$ is the outward unit normal vector of surface S. The electromagnetic fields were calculated using finite-difference time-domain (FDTD) simulation (Ansys Lumerical FDTD). Two Total Field Scattered Field (TFSF) sources, each linearly polarized in orthogonal directions and phase-shifted by π/2, were used to simulate the circularly polarized illumination in the $z$-direction. Six two-dimensional field monitors were positioned to form the faces of a 200 nm cubic surface enclosing the nanorod. The nanorod was modeled as a cylinder (25 nm diameter) with two hemispherical end caps, yielding a total length of 71 nm. Optical constants of gold were taken from Johnson and Christy.[10] The simulation volume inside the TFSF source (240 nm-sized cube) was uniformly meshed with a spatial resolution of 2 nm.

The resulting magnitude of optical torque is $\boldsymbol{\tau}$ = (0, 0, 2.0) pN nm per 1 mW µm$^{-2}$ of the illumination intensity at λ = 640 nm. And $\boldsymbol{\tau}$ = (0, 0, 2.1) pN nm per 1 mW µm$^{-2}$ of the illumination intensity at λ = 660 nm ($\tau_{opt,est}$ = *0.05 pN nm for a* $I_{inc,660}$ = 24 µW µm$^{-2}$, for the experimental conditions presented in Figure 5h).

## Supplementary Note 12: Layout of the cylindrical polarization-based interferometric scattering microscopes

We have used two configurations of the cypiSCAT microscope. The transmission arrangement (Supplementary Figure 9a), was used to observe surface-immobilized colloidal gold nanoparticles (Figure 3e-i) and track rotation of DNA origami-gold nanorod labels (Figures 4-6). The reflection arrangement (Supplementary Figure 9b) was used to measure the orientation of the gold nanorods patterned on a silicon wafer (Figure 3a-d).

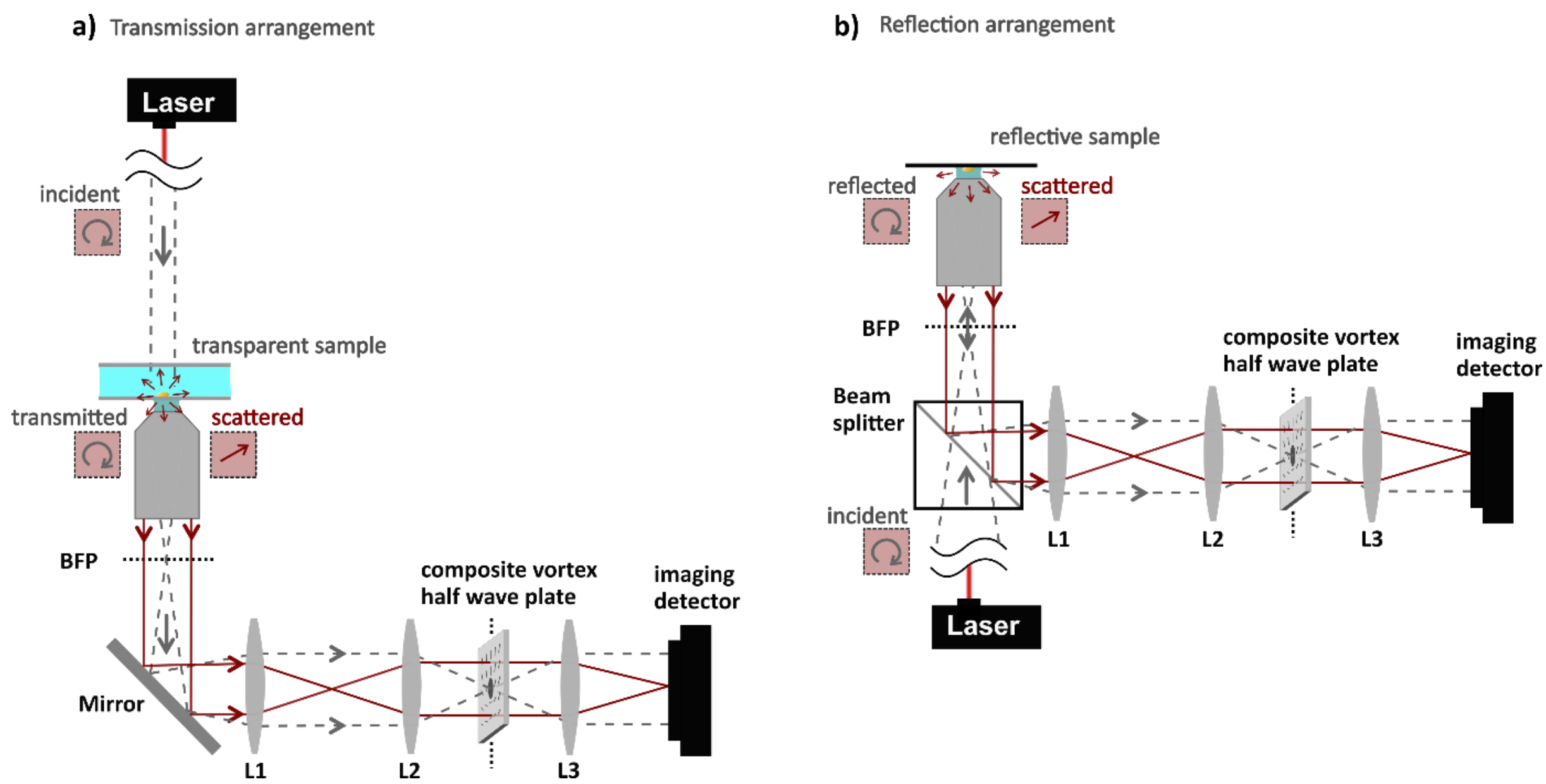

**Supplementary Figure 9** Simplified schematics of the cylindrical polarization-based iSCAT microscope. The polarization states (boxes with dotted edges) of the scattered wave are illustrated for the case of a highly anisotropic scatterer. **a)** The transmission arrangement, and **b)** the reflection arrangement.

## Supplementary Note 13: Preparation of the structured DNA orientation labels

Functional oligonucleotide strands were added to the structural DNA staple and scaffold strands (see Methods). These include six strands, denoted as NR-1 – NR-6 in Supplementary Table 4, designed to attach poly-T functionalized gold nanorod to the top part of the DNA origami, and a biotinylated strand, denoted as Rotor_BiotinExt1, to attach the rotor to the streptavidin-functionalized glass surface without constraining its rotational freedom.

| NR-1 | AATGCGCTAACGCCTGTGCTGCAAGGCGCCGCCTCTTCGCTAAAAAAAAAAAAAA |
|---|---|
| NR-2 | CACGTATTCTGCCATAGATGGGCGCATCTTGGCGGATTGACCAAAAAAAAAAAAA |
| NR-3 | AAACAGGTTCGCATTGTAAACGTTAATACTCAAAAACAGGAAAAAAAAAAAAAAA |
| NR-4 | GCCAGAAACCAGTAATGTACCGTAACACTACCTCATTTTCAGAAAAAAAAAAAAA |
| NR-5 | GAGTAAAAATTTATAAGACGCTGAGAAGCCCTTAGAATCCTTAAAAAAAAAAAAA |
| NR-6 | AATACTTAAGATGATTCATTTCAATTACGAGAATACCAAGTTAAAAAAAAAAAAA |
| Rotor_BiotinExt1 | /5Biosg/TTACACAACTGAGGAAAAAGGAACAACTAAA |

***Supplementary Table 4*** *Sequences of the oligonucleotide strands used for the attachment of the gold nanorod to the DNA origami ORBIT construct (NR-1 – NR-6) and the biotinylated strand used to attach the DNA origami to the surface without constraining the rotational motion.*

## Supplementary Note 14: Considerations for torque measurements based on rotation tracking of nanoscale objects in liquids

A torque acting on a nanoscopic object in a liquid environment is manifested by inducing the rotation of such an object. Nanoparticle in liquid operates in an overdamped regime, and the instant torque can be determined from the angular velocity $\omega$ and hydrodynamic rotational drag $\gamma_{rot}$ as $\tau = \omega\gamma_{rot}$. At the same time, the rotational motion is subject to random diffusive fluctuations due to thermal collisions with surrounding liquid molecules. To enable measurement of the deterministic torque-related dynamics over a random thermal fluctuations, the torque-related angular step has to be higher than the mean diffusive angular step over a time interval of the measurement $t$, i.e., $\Delta\alpha_{torque} > \langle\Delta\alpha_{diffusion}\rangle$.

Mean square angular displacement of a diffusive rotation is linearly related to the diffusion coefficient, in case of 1D rotation, $\langle\Delta\alpha^2_{diffusion}\rangle = 2D_{rot}t$ and the diffusion coefficient is related to the thermal energy of the surrounding medium as $D_{rot} = {k_BT}/{\gamma_{rot}}$ yielding: $\langle\Delta\alpha_{diffusion}\rangle = \sqrt{\frac{2k_BTt}{\gamma_{rot}}}$.

Assuming the spherical nanoparticle for simplicity, the rotational drag is $\gamma_{rot} = 8\pi\eta R^3$, $\eta$ is the viscosity of the medium and the condition $\Delta\alpha_{torque} = \omega t = \frac{\tau\, t}{\gamma_{rot}} > \langle\Delta\alpha_{diffusion}\rangle$ translates to:

$$\tau > \sqrt{\frac{16\pi\eta R^3 k_B T}{t}}. \tag{S16}$$

The right-hand side of Equation (S16) represents the diffusion-equivalent torque, i.e., the torque required to produce an angular displacement equal to the mean displacement caused by thermal diffusion over time $t$. Considering $T$ = 296 K, supplementary Figure 10a illustrates diffusion-equivalent torque as a function of nanoparticle radius for time intervals ranging from 1 µs to 1 s. Supplementary Figure 10b provides an alternative visualization, comparing torque-induced angular displacement ($\tau$=10pN nm, solid lines) with mean diffusive displacement (dashed lines) for spheres of radii 10, 100, and 1000 nm. To enable quantitative torque estimation from angular traces, the torque-induced displacement must exceed the diffusive displacement by a factor of $m$. This condition is met at a time $m^2$times longer than the diffusion-torque-equivalent time, corresponding to the crossover point of the solid and dashed curves in Supplementary Figure 10b.

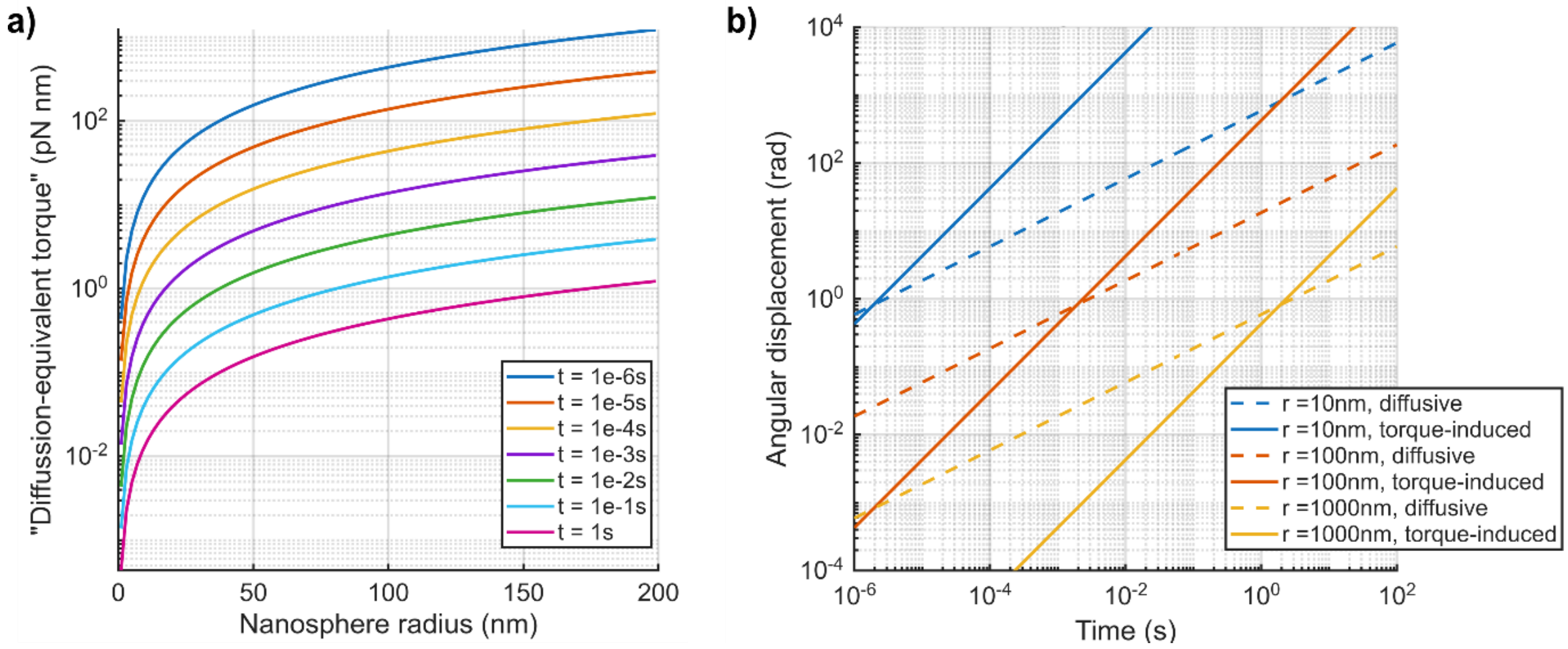

***Supplementary Figure 10*** *The limits of the nanoscale torque measurements.* ***a)*** *Diffusion-equivalent torque as a function of nanoparticle radius for various measurement time intervals ranging from 1 µs to 1 s.* ***b)*** *Comparison of torque-induced angular displacement (solid lines) and mean diffusive angular displacement (dashed lines) over time for spherical orientation labels of radii 10 nm, 100 nm, and 1000 nm. The crossover points indicate the diffusion-torque-equivalent time, beyond which deterministic torque-induced rotation dominates over thermal fluctuations.*

As an illustrative example, let's assume a torque τ = 10 pN nm acting on a spherical nanoparticle with a radius $r$ in water at $T$ = 296 K. According to Supplementary Figure 10, to observe 10-times larger torque-driven angular displacement compared to the mean diffusive angular displacement over a time $t_{exp}$, allowing for reliable quantification of the torque, the time interval required for the measurement is about 200 s, 200 ms, and 200 µs for spherical nanoparticles with r = 1000 nm, 100 nm, and 10 nm, respectively. As a rough estimate, to measure rapidly changing torques within a process such as enzymatic DNA unwinding, where the duration of the single reaction step may be below 1 ms, the size of the orientation label should be chosen to not exceed the hydrodynamic drag of a sphere with a radius of approximately 10 nm.

# Supplementary References